\documentclass[superscriptaddress, reprint, amsmath, amssymb, aps, prx, floatfix]{revtex4-1}
\usepackage{graphicx}
\usepackage{float}
\usepackage{wrapfig}
\newcommand{\norm}[1]{\left\lVert #1 \right\rVert}
\usepackage{dcolumn}
\usepackage{bm}
\usepackage{hyperref}
\usepackage{natbib}

\usepackage{xcolor}
\usepackage{soul}

\newcommand{\ie}{\emph{i.e.}, }
\newcommand{\eg}{\emph{e.g.}, }

\begin{document}

\title{Digital quantum simulation of molecular dynamics and control}

\author{Alicia B. Magann}
\email{amagann@princeton.edu}
\affiliation{Department of Chemical \& Biological Engineering, Princeton University, Princeton, New Jersey 08544, USA}
\affiliation{Extreme-Scale Data Science \& Analytics, Sandia National Laboratories, Livermore, California 94550, USA}

\author{Matthew D. Grace}
\email{mgrace@sandia.gov}
\affiliation{Extreme-Scale Data Science \& Analytics, Sandia National Laboratories, Livermore, California 94550, USA}

\author{Herschel A. Rabitz}
\email{hrabitz@princeton.edu}
\affiliation{Department of Chemistry, Princeton University, Princeton, New Jersey 08544, USA}

\author{Mohan Sarovar}
\email{mnsarov@sandia.gov}
\affiliation{Extreme-Scale Data Science \& Analytics, Sandia National Laboratories, Livermore, California 94550, USA}

\date{\today}

\begin{abstract}

Optimally-shaped electromagnetic fields have the capacity to coherently control the dynamics of quantum systems, and thus offer a promising means for controlling molecular transformations relevant to chemical, biological, and materials applications. Currently, advances in this area are hindered by the prohibitive cost of the quantum dynamics simulations needed to explore the principles and possibilities of molecular control. However, the emergence of nascent quantum-computing devices suggests that efficient simulations of quantum dynamics may be on the horizon. In this article, we study how quantum computers could be employed to design optimally-shaped fields to control molecular systems. We introduce a hybrid algorithm that utilizes a quantum computer for simulating the field-induced quantum dynamics of a molecular system in polynomial time, in combination with a classical optimization approach for updating the field. Qubit encoding methods relevant for molecular control problems are described, and procedures for simulating the quantum dynamics and obtaining the simulation results are discussed. Numerical illustrations are then presented that explicitly treat paradigmatic vibrational and rotational control problems, and also consider how optimally-shaped fields could be used to elucidate the mechanisms of energy transfer in light-harvesting complexes. Resource estimates, as well as a numerical assessment of the impact of hardware noise and the prospects of near-term hardware implementations, are provided for the latter task.

\end{abstract}

\maketitle
  
\section{Introduction}

\begin{figure}
\includegraphics[width=1\columnwidth]{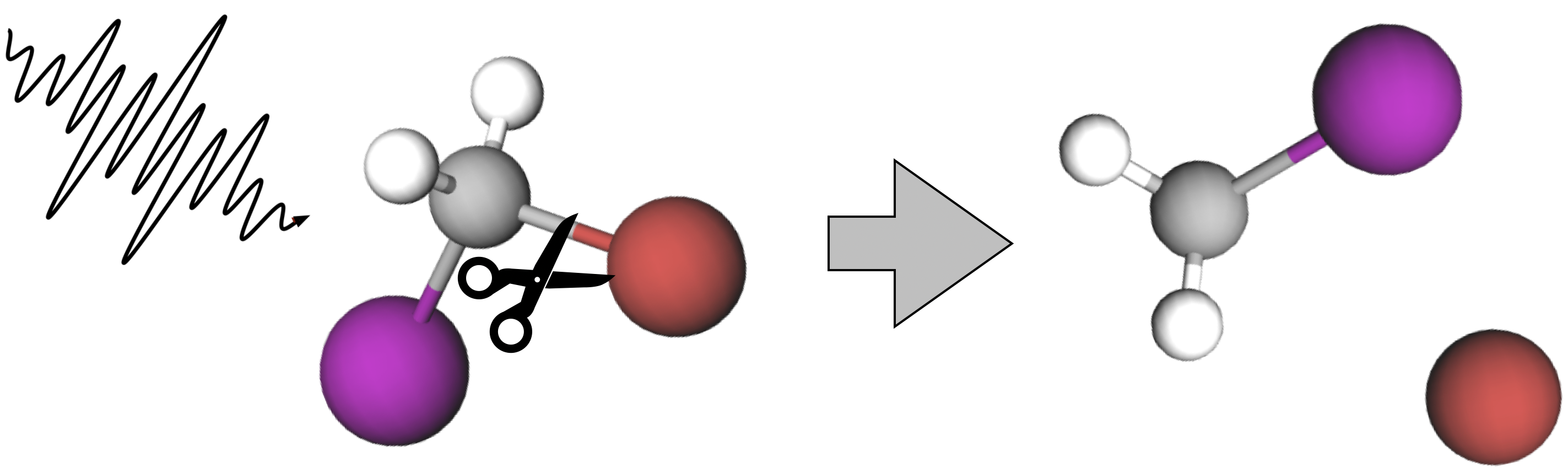}
\caption{The development of quantum optimal control theory was initially motivated by the goal of controlling selective dissociation reactions in molecules, as depicted, and it offered a flexible approach for realizing this goal. The development of femtosecond lasers and pulse-shaping technology in the 1990s subsequently provided the laboratory tools for the task, leading to several proof-of-principle demonstrations of control over selective dissociation using quantum optimal control \cite{Assion919,2002EPJD...20...71D,Levis709,Daniel536}. Today, quantum optimal control has found applications across chemical, biological, and materials applications \cite{Brif2011,Glaser2015}}
\label{QuantumControlFig}
\end{figure}

The ability to accurately control the dynamics of quantum systems would have significant implications across the physical sciences. Shaped electromagnetic fields able to coherently interact with molecules on their natural length and time scales offer an unprecedented tool for realizing such control, and there is growing interest in using them to control quantum systems with chemical, biological, and materials applications \cite{Brif2011,Glaser2015,doi:10.1021/acs.accounts.8b00244,RevModPhys.91.035005}. One method for designing shaped fields capable of steering a quantum system towards a desired control target is quantum optimal control, whose original development in the 1980s was driven by the dream of tailoring laser fields to control the outcomes of chemical reactions, as depicted in Fig.~\ref{QuantumControlFig}. Optimal control has since been realized in numerous proof-of-concept experiments involving the control of branching ratios of chemical reactions \cite{Assion919}, bond selective dissociation \cite{2002EPJD...20...71D,Levis709}, molecular fragmentation \cite{Daniel536}, bond making \cite{PhysRevLett.114.233003}, and isomerization in the liquid phase \cite{PhysRevLett.94.068305} as well as in the biologically-relevant retinal molecule bacteriorhodopsin \cite{Prokhorenko1257}. Laboratory quantum optimal control demonstrations have also spanned numerous applications beyond the control of chemical reactions, including the control of molecular alignment and orientation \cite{2009NatPh...5..289G}, decoherence mitigation in gas-phase molecules \cite{Branderhorst638}, preparation of coherent superposition states in molecules at room temperature \cite{2011NatPh...7..172H}, molecular optimal dynamic discrimination \cite{PhysRevLett.102.253001}, isotope selection \cite{PhysRevLett.93.033001}, high harmonic generation \cite{RevModPhys.80.117}, and energy flow in light-harvesting complexes \cite{Herek2002QuantumCO}. 

Despite the promise of these experimental demonstrations, quantum optimal control has not yet found wide, practical applications in molecular systems. A primary contributing factor is the lack of theoretical support. That is, the prohibitive computational costs associated with performing accurate quantum control simulations limit our ability to identify new quantum control applications, design new quantum control experiments, and assess the feasibility of achieving desired control outcomes in a given experimental setting. These costs also limit the quality of the analyses that can be performed to probe the control mechanisms underlying quantum control experiments. The challenges arise from the fact that the computational memory and time costs associated with simulating quantum dynamical systems without approximations scale exponentially in the number of degrees of freedom in the system, termed the ``curse of dimensionality.'' For molecular control simulations, this computational challenge is manifested in problems where the control of multiple coupled rotational, vibrational, and/or electronic degrees of freedom is sought.

A first solution to this challenge is to use a tractable, reduced model to simulate the field-induced quantum dynamics. Such models typically assume one or multiple approximations that can lead to deterioration of the solution accuracy, and while numerous approximate methods for quantum dynamics simulations have been developed, no method is suitable for every problem. For example, mean-field approaches such as time-dependent Hartree can be used to simulate controlled quantum molecular dynamics with costs polynomial in the system size, but these approaches often yield poor performance for systems with only a few degrees of freedom, or for systems whose degrees of freedom are strongly coupled \cite{Messina1996}. Improvements in accuracy can be gained by using variants such as multi-configurational time-dependent Hartree, but these alternatives can have costs that scale exponentially in the system size and are therefore not suitable for large systems \cite{Schroder2008}. For simulations of controlled multi-electron dynamics, time-dependent Hartree-Fock \cite{Mundt_2009} and its variants can be used, but suffer from the same drawbacks. Alternatively, time-dependent density functional theory \cite{PhysRevLett.109.153603} can be used, but the choice of exchange-correlation functional yields approximations that are not well understood. Other approaches include tensor network methods such as the time-dependent density matrix renormalization group \cite{PhysRevLett.106.190501}, which become prohibitively expensive as the simulation time increases.

A second solution is to seek alternative simulation settings that do not suffer from this exponential scaling problem: analog or digital quantum simulators show promise for this purpose \cite{RevModPhys.86.153}. Analog quantum simulation involves tuning a controllable quantum system such that it emulates the behavior or dynamics of another specific quantum system of interest. Analog quantum simulators could have high value for specific applications, and the use of analog quantum simulators for quantum control simulations has been explored recently. For example, in ref.~\cite{PhysRevLett.118.150503}, fields for preparing a particular correlated spin state were optimized using a gradient algorithm on an NMR quantum simulator, and in refs.~\cite{2019arXiv191100789D,2020arXiv200201068Y}, the design of bang-bang control protocols for one- and multi-qubit systems using noisy quantum devices was considered. Meanwhile, in ref.~\cite{2018Natur.557..660S}, quantum photonics were used for the analog simulation of quantum vibrational dynamics and control, where the initial state of ammonia was optimized to maximize the probability of molecular dissociation. However, due to their analog nature, there are concerns about the reliability of the solutions that such simulators produce when scaled to large systems \cite{hauke_can_2012, sarovar_reliability_2016}. 

A universal alternative to analog quantum simulation is digital quantum simulation, which can be used to simulate the dynamics of general quantum systems through a set of discrete operations. The most common model for digital quantum simulators is the circuit-model of quantum computation, which can simulate arbitrary unitary evolutions using operations in the form of quantum circuits \cite{Nielsen2010}. Research to develop circuit-model quantum computers has accelerated in recent years, and technologies based on superconducting qubits \cite{Wendin_2017} and trapped ions \cite{doi:10.1063/1.5088164} capable of implementing shallow quantum circuits on tens of qubits are currently available. To sustain this progress, there is a need to explore scientifically-relevant problems for which quantum computers offer clear advantages over their classical counterparts, and to define candidate problems for evaluating their performance.

In this article, we explore how a quantum computer could be used as a digital quantum simulator for the design of quantum optimal controls for molecular systems. To this end, we introduce a hybrid quantum-classical scheme combining (a) digital quantum simulation methods for simulating the molecular dynamics in polynomial time \cite{10.2307/2899535} with (b) classical optimization approaches to identify control fields for achieving a desired task. Our scheme offers a clear example of a scientific problem amenable to a quantum speedup, which we hope will serve to motivate current efforts advancing quantum computing devices. The first step of this scheme involves encoding the state and Hamiltonian of the molecular system under consideration into qubits for simulation on the quantum computer. To date, relatively little attention has been given to encoding procedures for simulating rotational and vibrational systems \cite{2018arXiv181104069M, 2019arXiv190912847S}, which are common and important applications for quantum control. We address this by outlining a general encoding approach, and provide explicit details regarding applications to rotational and vibrational systems.

We also consider the applicability of this general scheme towards elucidating the mechanisms underpinning important light-matter interactions found in nature, such as the absorption of sunlight and transport of photoexcitations by pigments in light-harvesting complexes of photosynthetic organisms. This process marks the first stage of photosynthesis and is widely believed to involve quantum coherent excitonic dynamics at short time-scales \cite{2005Natur.434..625B,2007Natur.446..782E, scholes_quantum-coherent_2010, scholes_solar_2012}. Numerical studies are needed for understanding this process, but require simulating the quantum dynamics of numerous coupled pigments interacting with a larger, thermal environment, which poses a significant computational challenge  \cite{strumpfer_open_2012, kreisbeck_scalable_2014, 2020arXiv200103685B}. The mechanisms underlying photosynthesis can also be probed using two-dimensional electronic spectroscopy experiments, which produce maps of the energy transfer in light-harvesting complexes after a particular initial electronic excitation has been prepared \cite{schlau-cohen_two-dimensional_2011,doi:10.1146/annurev-conmatphys-020911-125126}. Thus, in this latter setting, the ability to prepare the complex in a state that leads to the desired energy transfer dynamics is of paramount importance. We consider how optimally-shaped fields could be used to this effect, and estimate the qubit counts and circuit depths needed to perform the associated simulations on a quantum computer. Using this setting, we also consider the prospects of implementing of our approach on current and near-term quantum hardware. In particular, we numerically analyze the performance of our algorithm in the presence of different levels of noise, using a model for quantum hardware based on trapped ions.

The remainder of the article is organized as follows. Section \ref{Sec:Quantum optimal control theory} provides an overview of quantum optimal control theory and the computational challenge associated with simulating the dynamics of molecular systems with multiple degrees of freedom. In Section \ref{Sec:Hybrid algorithm for quantum optimal control}, we introduce a hybrid quantum algorithm that eliminates the computational challenge by performing the quantum dynamics simulation on a quantum computer, which can perform the simulation in polynomial time, while using a classical co-processor only to store and update the coefficients parameterizing the control field. Details regarding digital quantum simulation are given in Section \ref{Sec:Digital quantum simulation}, with the qubit encoding, Hamiltonian simulation, and qubit readout each discussed in subsections \ref{Sec:Qubit encoding}, \ref{Sec:Hamiltonian simulation}, and \ref{Sec:Qubit readout}, respectively. Details regarding the classical optimization are given in Section \ref{Sec:Classical optimization}. A series of numerical illustrations are presented in Section \ref{Sec:Numerical illustrations}. First, an illustration involving the control of bond stretching in hydrogen fluoride is described in Section \ref{Sec:Controlled bond stretching in diatomic molecule}, followed by illustrations involving the controlled orientation of dipole-dipole coupled molecular rotors in Section \ref{Sec:Controlled orientation of two dipole-dipole   coupled molecular rotors}, the controlled state preparation in a light-harvesting complex in Section \ref{Sec:Controlled state preparation in   light-harvesting complex}, and a numerical analysis of the impacts of hardware noise on the algorithm performance in Section \ref{Sec:Effects of hardware noise on algorithm performance}. We conclude with a look to future research directions in Section \ref{Sec:Outlook}.

\section{Quantum optimal control theory} \label{Sec:Quantum optimal control theory}

An example of a general quantum optimal control problem relevant to this work consists of designing a control field $f(t,\{\theta_i\})$ for $t\in[0,T]$, parameterized by a set of coefficients $\theta_i$, $i = 1,\cdots,K$, that will achieve a desired control objective at the terminal time $t = T$. This can be posed as a minimization problem:
\begin{equation}
\begin{aligned}
\min_{\{\theta_i\}} J[T,\{\theta_i\}]\, ,
\end{aligned}
\label{ObjFunc}
\end{equation}
where $J[T,\{\theta_i\}]$ is the objective function, which is formulated to include the control target, often with additional criteria, which can be defined to reflect the resources available in an associated laboratory implementation. One common choice is to seek low-energy fields for achieving a particular control target by including a term penalizing the field fluence \cite{PhysRevA.37.4950}. For problems involving multiple control fields, the search is performed with respect to the set of coefficients $\{\theta_i\}$ which, taken together, parameterize the set/space of available controls. In vibrational control problems, the set $\{\theta_i\}$ often contains the amplitudes and phases of the frequency components of the laser field. For control problems involving longer time-scales, the pulse can often be modulated in the laboratory directly in the time-domain using an arbitrary waveform generator \cite{Lin2005,Yao2011}. 

Quantum optimal control simulations are chiefly useful for identifying new quantum control applications, analyzing the feasibility of controlling new classes of quantum phenomena, and providing a basic understanding of controlled quantum dynamics. When numerically-designed control fields are applied to actual molecular systems in the laboratory, however, a significant loss of fidelity can occur. This can happen due to noise-based fluctuations in the applied field, uncontrolled interactions with the environment, and uncertainties in the molecular Hamiltonian, including uncertainties in the description of the molecular dipole moment, which couples the field and the molecular system. Thus, it is important to identify fields that are robust to such errors and uncertainties if fields are sought for a direct laboratory implementation. This can be accomplished by including additional robustness criteria in $J[T,\{\theta_i\}]$ \cite{PhysRevA.49.2241,PhysRevA.71.063806}, or by designing fields based on modeling uncertainties using a statistical distribution \cite{PhysRevA.42.1065}. 

In simulations, the optimal control field parameters $\{\theta_i\}$ that minimize $J[T,\{\theta_i\}]$ are often sought iteratively. To evaluate $J[T,\{\theta_i\}]$ at each iteration, the dynamics of the quantum system under consideration, driven by the field $f(t,\{\theta_i\})$ with a particular parameterization $\{\theta_i\}$, must be simulated by solving the time-dependent Schr\"odinger equation:
\begin{equation}
i\frac{\partial}{\partial t}|\psi(t)\rangle = H(t,\{\theta_i\})|\psi(t)\rangle\,,
\label{TDSE}
\end{equation}
where $|\psi(t)\rangle$ is the system state at time $t$ (we set $\hbar = 1$). 

The Hamiltonian $H(t,\{\theta_i\})$ of the system is assumed to be ``$k$-local'', \ie to contain interactions coupling up to $k$ degrees of freedom (these are sometimes also called ``$k$-body'' interactions), for a constant $k$ that does not scale with the total number of degrees of freedom $M$. In models of low-energy physics, \eg those derived from quantum electrodynamics (QED), interactions are typically limited to two-body terms, such as the Coulomb interaction derived from the minimal coupling Hamiltonian in QED. There are some exotic settings where $k$-body interaction terms with $k>2$ are present \cite{murphy_three-body_1971}, but these are weak in comparison to the $k=2$ terms, and moreover, $k$ never scales with $M$, the number of elementary degrees of freedom. Thus, the $k$-local assumption is not a strong one. 
In general, $H(t,\{\theta_i\})$ can be expressed in the dipole approximation as
\begin{equation}
H(t,\{\theta_i\})=H_0+H_cf(t,\{\theta_i\})\,,
\label{TotalHamiltonian}
\end{equation}
where $H_0$ is the time-independent molecular ``drift'' Hamiltonian, which contains all kinetic and field-free potential terms, including potentials due to fixed interactions between various degrees of freedom. The control Hamiltonian $H_c$ describes the light-matter interaction underlying the coupling of the molecular dipole moment of the system to the applied field $f(t,\{\theta_i\})$. In cases where multiple fields are applied, Eq. (\ref{TotalHamiltonian}) also includes additional terms describing the coupling of each field to the system. 

Frontier applications of quantum optimal control often involve complex quantum molecular systems with multiple interacting degrees of freedom. The Hilbert space dimension of such systems scales exponentially in the number of degrees of freedom present, leading to an explosion of the computational resources required for simulating the system dynamics, rapidly rendering quantum optimal control simulations intractable. As such, despite the breadth of theoretical research on quantum optimal control theory, applications of quantum optimal control to molecular systems with many degrees of freedom remain scarce.

\section{Hybrid algorithm for quantum optimal control} \label{Sec:Hybrid algorithm for quantum optimal control}

\begin{figure*}
\includegraphics[width=1.7\columnwidth]{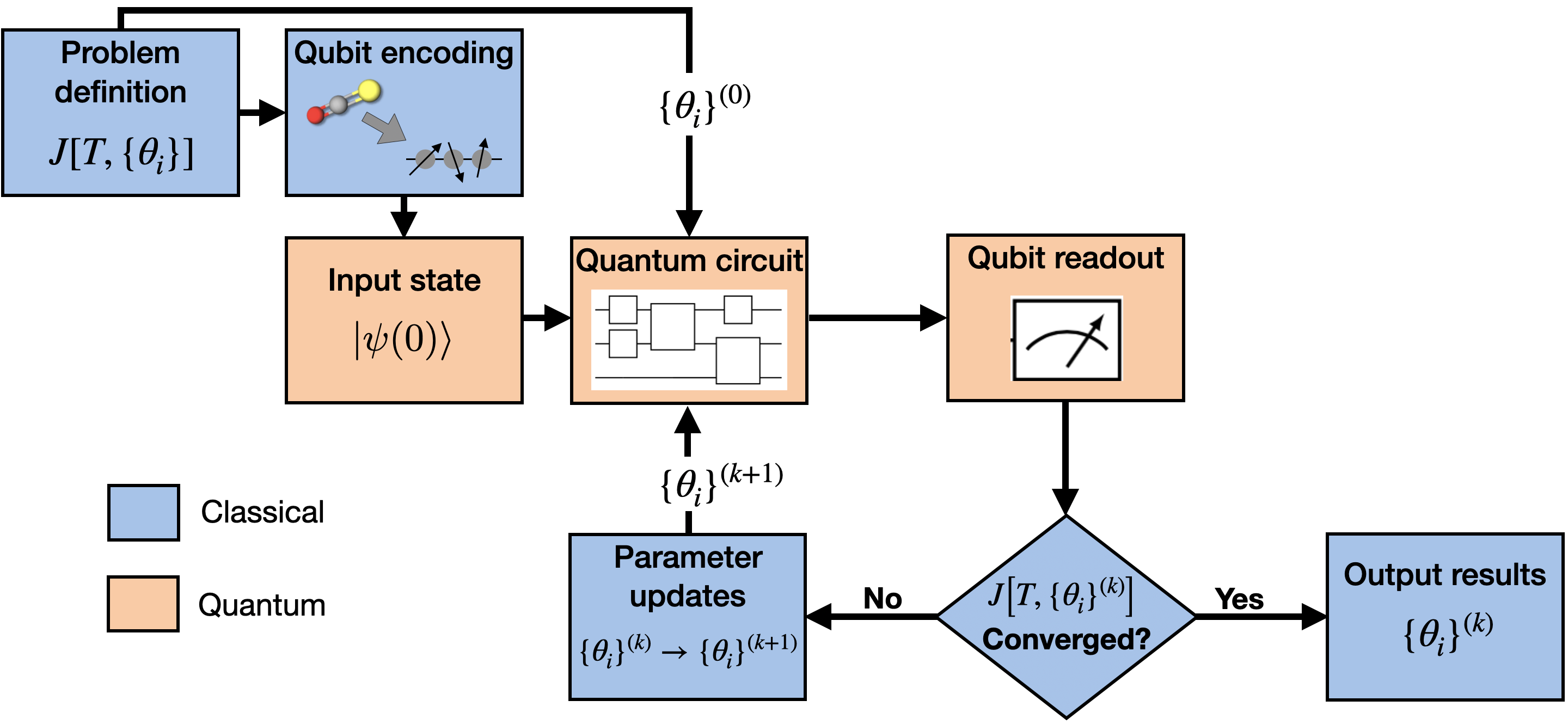}
\caption{Diagram of hybrid quantum-classical simulation algorithm for designing optimal controls for molecular systems. At the outset, a molecular quantum control problem is defined and encoded into qubits. Next, the qubits are initialized in a state $|\psi(0)\rangle$, representing the encoding of the state of the molecular system at time $t=0$. A quantum circuit is then used to simulate the dynamics of the quantum system driven by a control field, parameterized by the set of coefficients $\{\theta_i\}$,  up to the terminal time $t=T$. Then, the qubits are measured to obtain the value of the control objective function $J[T,\{\theta_i\}]$. If $J[T,\{\theta_i\}]$ has converged to a desired tolerance, the parameters $\{\theta_i\}$ are returned as the results of the simulation, if not, a classical co-processor is used to update the set $\{\theta_i\}$, which are then returned to the quantum computer for subsequent iteration.}
\label{SimulationFrameworkDiagram}
\end{figure*}

A common goal of quantum optimal control simulations is the identification of a set of parameters $\{\theta_i\}$ describing a control field $f(t,\{\theta_i\})$, $t\in[0,T]$, that achieves a specified objective as well as possible at the terminal time $T$. Quantum optimal control simulations thus have two key components: the evaluation of the control objective function $ J[T,\{\theta_i\}]$ for a particular set of control parameters $\{\theta_i\}$, and the updates of $\{\theta_i\}$ according to a chosen optimization algorithm. We propose a hybrid quantum-classical scheme for these simulations, where a quantum computer is used for efficiently evaluating $ J[T,\{\theta_i\}]$ by simulating the driven dynamics of a quantum system, while a classical coprocessor is used for updating the control parameters $\{\theta_i\}$\footnote{Quantum optimization algorithms could also be considered for this purpose; however, given the lack of a demonstrated advantage of using quantum hardware for the optimization purposes considered here, we do not explore this possibility in this work.}, as depicted in Fig.~\ref{SimulationFrameworkDiagram}. 

\subsection{Digital quantum simulation} \label{Sec:Digital quantum simulation}

\subsubsection{Qubit encoding} \label{Sec:Qubit encoding}

The initial task associated with simulating quantum dynamical systems is the choice of a finite representation for the system state and Hamiltonian. This is true for simulations on both quantum and classical hardware, as only finite computational resources are available in both settings. For continuous-variable systems such as molecules, whose Hilbert spaces are inherently infinite dimensional, simulations must be performed in a suitable truncated space. One approach is to represent the system state and associated operators in real space, using a finite mesh with differential operators represented using finite differences. Another approach is to represent the system state and associated operators using a particular finite set of basis functions, which are often chosen to be orthonormal.

After a finite representation is chosen, the corresponding quantum control problem must be encoded into qubits for implementation on a quantum computer. A variety of encodings have been developed for this purpose, e.g., refs.~\cite{2018arXiv180810402M,2019arXiv190912847S}; here, we focus on general basis set encodings that are relevant to quantum control problems \footnote{For details regarding real space encodings, which treat the dynamics of nuclei and electrons equally, and may therefore be preferable for simulations of molecular dynamics outside of the Born-Oppenheimer approximation, we refer the interested reader to ref. \cite{Kassal18681}.}. For basis set encodings, the choice of basis set affects the cost of the initial qubit state preparation, the circuit depth and width required to simulate the dynamics of the molecular system, and the complexity of obtaining the value of $J[T,\{\theta_i\}]$ at the terminal time. Consequently, the basis could be chosen with the goal of balancing all of these costs, or, if one task is particularly challenging, the basis could instead be chosen to minimize the complexity of this particular task. 

In a basis set encoding, a set of $d$ basis states $\{|q\rangle\}$, $q = 1,2,\cdots,d$, can be mapped to qubit states as
\begin{equation}
\begin{aligned}
    |1\rangle& \mapsto\bar{|1\rangle}=|0\rangle|0\rangle\cdots|0\rangle|0\rangle|0\rangle\\
    |2\rangle& \mapsto\bar{|2\rangle}=|0\rangle|0\rangle\cdots|0\rangle|0\rangle|1\rangle\\
    |3\rangle& \mapsto\bar{|3\rangle}=|0\rangle|0\rangle\cdots|0\rangle|1\rangle|0\rangle\\
    |4\rangle& \mapsto\bar{|4\rangle}=|0\rangle|0\rangle\cdots|0\rangle|1\rangle|1\rangle\\
    |5\rangle& \mapsto\bar{|5\rangle}=|0\rangle|0\rangle\cdots|1\rangle|0\rangle|0\rangle,\,\text{etc.,}\\
    \end{aligned}
\end{equation}
using a standard binary mapping, while an arbitrary state can be represented as a superposition of $d$ basis states as $|\psi(t)\rangle = \sum_{q=1}^{d}c(q,t)|q\rangle$, where $c(q,t)$ is the probability amplitude associated with the basis state $|q\rangle$ at time $t$. In this manner, basis set encodings can be used to encode the state of a quantum degree of freedom represented with $d$ basis states in $\lceil\log_2 d\rceil$ qubits, where $\lceil \cdot \rceil$ is the ceiling function. The full $2^{\lceil\log_2 d\rceil}$ dimensional space associated with each degree of freedom is then spanned by the Pauli operator basis $\{B_\ell\}_{\ell=1}^{2^{2\lceil\log_2 d\rceil}}$, where each 
\begin{equation}
    B_\ell = \bigotimes_{s=1}^{\lceil\log_2 d\rceil} N_\sigma \sigma_{s}^{(\ell)}
\end{equation}
is a (normalized) Pauli string, where $N_\sigma = 1/\sqrt{2}$ is a prefactor included for normalization, and $\sigma$ denotes one of the Pauli operators $\sigma_x = \bigl( \begin{smallmatrix}0 & 1\\ 1 & 0\end{smallmatrix}\bigr)$, $\sigma_y = \bigl( \begin{smallmatrix}0 & -i\\ i & 0\end{smallmatrix}\bigr)$, $\sigma_z = \bigl( \begin{smallmatrix}1 & 0\\ 0 & -1\end{smallmatrix}\bigr)$, or $\sigma_I = \bigl( \begin{smallmatrix}1 & 0\\ 0 & 1\end{smallmatrix}\bigr)$ on qubit $s$. Thus, any operator $A$ acting on the degree of freedom can be encoded into a weighted sum of Pauli strings by projecting it onto the Pauli basis as
\begin{equation}
    A \mapsto \bar{A} = \sum_{\ell=1}^{2^{2\lceil\log_2 d\rceil}} g_\ell B_\ell\,,
    \label{OperatorEncoding1}
\end{equation}
where $\bar{A}$ denotes the encoded version of the operator $A$ and the coefficients $g_\ell= \langle A, B_\ell\rangle_{\textrm{HS}}$ can be computed from the Hilbert-Schmidt inner product between $A$ and each of the Pauli basis operators $B_\ell$. If $d$ is not an exact power of 2, then prior to the encoding, the $d\times d$ dimensional matrix $A$ should be expanded to $2^{\lceil\log_2 d\rceil}\times 2^{\lceil\log_2 d\rceil}$ dimensions by adding zeros.

For molecular control problems, we are primarily concerned with qubit encodings relevant to systems consisting of multiple coupled degrees of freedom. The framework outlined above can be straightforwardly generalized to such cases. Namely, for quantum systems with $M$ coupled degrees of freedom, each represented using $d$ basis states, 
\begin{equation}
    N=M\lceil\log_2 d\rceil
    \label{NumQubits}
\end{equation}
qubits can be used to represent the full system state in the associated $2^N$ dimensional Hilbert space as $|\psi(t)\rangle = \sum_{q_1,\cdots,q_M}c(q_1,\cdots,q_M,t)|q_1,\cdots,q_M\rangle$, using $M$ registers containing $\lceil\log_2 d\rceil$ qubits each. Eq. (\ref{NumQubits}) states that the number of qubits $N$ required to represent $|\psi(t)\rangle$ on a quantum computer scales linearly in the number of degrees of freedom $M$. This can be contrasted with the memory resources needed to represent $|\psi(t)\rangle$ on a classical computer, which scale exponentially as $d^M$. 

The operator encoding for the associated $k$-local Hamiltonian can be performed as
\begin{equation}
H(t) \mapsto \bar{H}(t) = \sum_{\ell=1}^{L} g_\ell(t) B_\ell
\label{OperatorEncoding}
\end{equation}
according to Eq. (\ref{OperatorEncoding1}), where the number of Pauli strings $L$ in the decomposition has the upper-bound
\begin{equation}
    L \leq \frac{M! }{k!(M-k)!} \,2^{2k\lceil\log_2 d\rceil} ,
    \label{Mchoosek}
\end{equation}
and where each Pauli string acts nontrivially on a $2^{k\lceil\log_2 d\rceil}$ dimensional space (for further details, see Appendix \ref{Sec:Appendix}). These $L$ operators can be computed by decomposing each local term in the Hamiltonian classically, which requires the classical resources to store and manipulate the associated $2^{k\lceil\log_2 d\rceil}\times 2^{k\lceil\log_2 d\rceil}$ dimensional matrices. For systems with multiple electronic degrees of freedom, Fermi statistics must also be enforced (\eg by using the Jordan-Wigner, parity, or Bravyi-Kitaev mappings, which automatically enforce Fermi statistics at the operator level \cite{2018arXiv180810402M}). 

\subsubsection{Hamiltonian simulation} \label{Sec:Hamiltonian simulation}

At the outset, the qubits must be prepared in the state $|\psi(0)\rangle$ encoding the initial condition of the molecular system. Then, the system's time evolution can be simulated by applying a quantum circuit to approximate the quantum time evolution operator $U(T,0)$, defined as the solution to Eq. (\ref{TDSE}), given by $U(T,0) = \mathcal{T}e^{-i\int_0^TH(t)dt}$, where $\mathcal{T}$ denotes the time-ordering operator. The control field is assumed to be piecewise-constant over a sequence of $N_t$ time steps of length $\Delta t$, and consequently, $U(T,0)$ can be computed as the time-ordered product $ U(T,0) = U(T,T-\Delta t)\cdots U(2\Delta t, \Delta t)\,U(\Delta t,0)$, where each term in the product is generated by a time-independent Hamiltonian and can be approximated using product formulas in polynomial time \cite{10.2307/2899535}. 

After the actual molecular Hamiltonian has been encoded as $\bar{H}$, it is expressed as a weighted sum of Pauli strings according to Eq.~\eqref{OperatorEncoding}, where for $k$-local Hamiltonians, $L$ grows polynomially with the number of qubits $N$ as per Eq. (\ref{Mchoosek}). Then, the first-order product formula is given by 
\begin{equation}
    U_{\textrm{PF}1}(t+\Delta t, t) =\prod_{\ell=1}^L \Big(e^{-ig_\ell(t) B_\ell\Delta t/n}\Big)^n\,,
    \label{PF1}
\end{equation}
where $n$ is the so-called Trotter number, which defines the accuracy of the approximation (\ie for $n\rightarrow\infty$, the first-order product formula is exact) \cite{10.2307/2899535}. The error incurred from using the first-order product formula is given by
\begin{equation}
    \epsilon_{\textrm{PF}1}(t+\Delta t,t) = \mathcal{O} \bigg(\frac{L^2\Lambda(t)^2 \Delta t^2}{n}\bigg)\,,
\end{equation}
where $\epsilon_{\textrm{PF}(k)}(t_2,t_1) = \norm{U(t_2,t_1)-U_{\textrm{PF}(k)}(t_2,t_1)}$, $\Lambda(t) = \max_\ell g_\ell (t)$, and $\norm{\cdot}$ denotes the spectral norm \cite{Childs9456}.  The total error $\epsilon_{\textrm{PF}1}(T,0)$ can be bounded using the triangle inequality by the sum of errors over each time-step \cite{Nielsen2010}, to yield
\begin{equation}
    \epsilon_{\textrm{PF}1}(T,0)= \mathcal{O} \bigg(\frac{N_t L^2\Lambda_{\max}^2 \Delta t^2}{n}\bigg)\,,
    \label{epsilonPF1}
\end{equation}
where $\Lambda_{\max} = \max_j\Lambda(t_j)$ and $N_t = T/\Delta t$. Higher-order product formulas can be used to improve the accuracy of the approximation \cite{doi:10.1063/1.529425}, and can be defined recursively as
\begin{equation}
\begin{aligned}
    S_{2p}(t,\lambda)&=\Big(S_{2p-2}(t,\gamma_p\lambda)\Big)^2S_{2p-2}\big(t,(1-4\gamma_p)\lambda\big)\\
    &\qquad\times \Big(S_{2p-2}(\gamma_p\lambda)\Big)^2\,,
\end{aligned}
\end{equation}
where $\gamma_p = \left( 4-4^{1/(2p-1)} \right)^{-1}$, which can be seeded with
\begin{equation}
    S_2(t,\lambda)=\Big(\prod_{\ell=1}^Le^{g_\ell(t) B_\ell\lambda/2}\Big)\Big(\prod_{\ell=L}^1 e^{g_\ell(t) B_\ell\lambda/2}\Big)\,,
\end{equation}
such that 
\begin{equation}
    \begin{aligned}
    U_{\textrm{PF}2}(t+\Delta t,t)&= \Big(S_2(t,\tfrac{-i\Delta t}{n})\Big)^n 
    \end{aligned}
\end{equation} 
and
\begin{equation}
    \begin{aligned}
    U_{\textrm{PF}(2p)}(t+\Delta t,t) &=\Big(S_{2p}(t,\tfrac{-i\Delta t}{n})\Big)^n\,.
    \end{aligned}
\end{equation} 
The total error associated with $2p$th-order product formulas is \cite{Berry2007}
\begin{equation}
    \epsilon_{\textrm{PF}(2p)}(T,0)= \mathcal{O} \bigg(\frac{N_t(2L5^{p-1}\Lambda_{\max} \Delta t)^{2p+1}}{n^{2p}}\bigg)\,.
    \label{epsilonPF2p}
\end{equation}

The cost of approximating $U(T,0)$ on a quantum computer can be quantified by the number of qubits (\ie memory) and the circuit depth (\ie run time) required. Product formulas require no ancilla qubits, and so the number of qubits $N$ needed is the same as the number of qubits required for encoding the state and Hamiltonian, and is given in Eq. (\ref{NumQubits}). This stands in contrast to other quantum algorithms developed for simulating the time evolution of quantum systems, such as the Taylor series algorithm \cite{PhysRevLett.114.090502} and algorithms based on quantum walks, such as the quantum signal processing algorithm \cite{2016arXiv161006546H,PhysRevLett.118.010501}, which offer improved error scaling, but each require additional ancilla qubits. As such, product formulas may have greater utility for early devices with limited qubit counts. The asymptotic scaling of the circuit depth $D$, quantified by the number of applications of $e^{-i B_\ell\tau_\ell}$, for arbitrary $\tau_\ell$, is 
\begin{equation}
    D_{\textrm{PF}1} = \mathcal{O} \bigg( \frac{N_t L^3\Lambda_{\max}^2 \Delta t^2}{\epsilon}\bigg)
    \label{DPF1}
\end{equation} 
for the first-order product formula and
\begin{equation}
    D_{\textrm{PF}(2p)}=\mathcal{O} \bigg(\frac{5^{2p} N_t L(L\Lambda_{\max}\Delta t)^{1+1/2p}}{\epsilon^{1/2p}}\bigg)
    \label{DPF2p}
\end{equation}
for $2p$th-order product formulas \cite{Berry2007}. The expressions given in Eqs. (\ref{epsilonPF1}), (\ref{epsilonPF2p}), (\ref{DPF1}), and (\ref{DPF2p}) are known to be very loose, and consequently, the circuit depths required in practice to achieve an error bounded by some $\epsilon$ can be expected to be far lower (e.g., orders of magnitude lower \cite{Childs9456}) than the depths given by Eqs. (\ref{DPF1}) and (\ref{DPF2p}).

\begin{figure*}
\includegraphics[width=2.08\columnwidth]{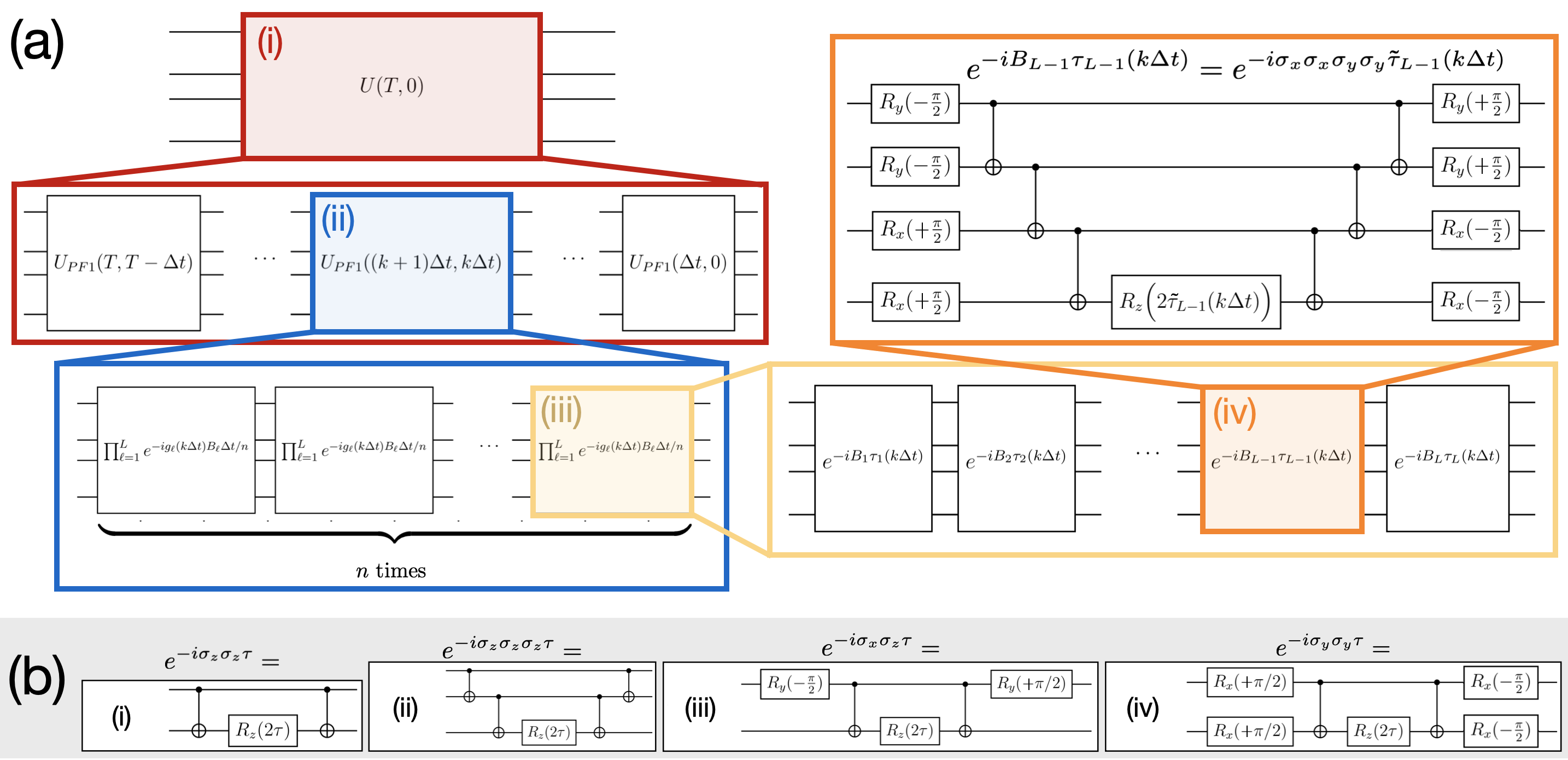} 
\caption{(a) Sample circuit diagram for implementing $U(T,0)$ on $N=4$ qubits using a first-order product formula. In (i), the time evolution operator $U(T,0)$ is decomposed as a product of piecewise-constant small time-step operators over each step of length $\Delta t$. Each small time-step operator can then be decomposed, as shown in (ii), into $n$ applications of $\prod_{\ell=1}^Le^{-ig_\ell(k\Delta t) B_\ell\Delta t/n}$, where each application can be further decomposed into a product of exponentials of Pauli strings, $e^{-iB_\ell\tau_\ell(k\Delta t)}$, as shown in (iii). Finally, each $e^{-iB_\ell\tau_\ell(k\Delta t)}$ can be implemented with a basic circuit on $\mathcal{O} (N)$ qubits; a sample circuit for $B_\ell=N_\sigma^4 \sigma_x\sigma_x\sigma_y\sigma_y$ is presented in (iv), where $\tilde{\tau}_{L-1}(k\Delta t)$ contains the normalization prefactor. \\ (b) Details on composing the basic circuit structure $ e^{-i B_\ell\tau_\ell(t)}$ from CNOT gates and one-qubit rotations $R_\sigma(\theta)\equiv e^{-i\sigma\theta/2}$. In (i), the basic circuit associated with $\otimes _{k=1}^1 \sigma_k^{(l)} = \sigma_z\sigma_z$ for $N=2$ qubits is shown, while (ii) shows how to generalize this circuit structure to additional qubits. Circuits (iii) and (iv) illustrate how to account for $\sigma_x$ and $\sigma_y$ terms, by adding one-qubit rotations that transform the $\sigma_z$ operations to $\sigma_x$ or $\sigma_y$ operations \cite{tacchino2020quantum}. These structures can be straightforwardly generalized to Pauli strings on any number of qubits, with any combination of Pauli operators \cite{Raeisi_2012}. }
\label{CircuitSchematic}
\end{figure*}

General quantum circuits for approximating $U(T,0)$ using product formulas can be designed using the observation that a quantum circuit able to implement $e^{-i B_\ell \tau_\ell(t)}$ for arbitrary scalar $\tau_\ell(t)$ is sufficient, as the full quantum algorithm can be constructed as a simple concatenation of circuits with this basic structure. Fig.~\ref{CircuitSchematic} illustrates how these basic circuits can be formed, where for $N$ qubits, the associated circuit depth required scales as $\mathcal{O} (N)$. Although the procedure outlined in Fig.~\ref{CircuitSchematic} can always be used to form quantum circuits to implement the algorithm, and Eqs. (\ref{DPF1}) and (\ref{DPF2p}) are useful for determining general bounds on circuit depth, significant gains can often be realized by using quantum compilers that seek to minimize the dominant costs (\eg two-qubit gates for noisy devices, or T-gates for error-corrected devices) when translating product formula algorithms into quantum circuits for implementation on particular hardware.

\subsubsection{Qubit readout} \label{Sec:Qubit readout}

After implementing the quantum circuit, the value of $J[T,\{\theta_i\}]$ is obtained by measuring the qubits. This can be performed efficiently provided that the operators in the objective function $J[T,\{\theta_i\}]$ whose expectation values are sought can be mapped to a set of $\text{poly}(N)$ qubit operators as per Section \ref{Sec:Qubit encoding}. Then, the expectation values of these operators can then be obtained by performing simultaneous projective measurements of the qubits in the computational $\sigma_z$ basis, and averaging over many runs. To measure terms in $J[T,\{\theta_i\}]$ containing $\sigma_x$ or $\sigma_y$, one-qubit rotations can be applied to transform the computational basis into the desired $\sigma_x$ or $\sigma_y$ basis prior to measurements \cite{2019arXiv190703505T,2020arXiv200103685B}, and, similarly, to measure the expectation values of multiqubit operators such as $\sigma_x\sigma_y\sigma_z$, the results of one-qubit measurements (in the appropriate rotated bases) can be multiplied together.

When the error $\epsilon_{\textrm{PF}(k)}(T,0)$ associated with using product formulas to simulate the dynamics is sufficiently small, then measurements performed on the state $|\psi_{\textrm{PF}(k)}(T)\rangle = U_{\textrm{PF}(k)}(T,0)|\psi(0)\rangle$ in order to determine the value of $J[T,\{\theta_i\}]$ are guaranteed to yield approximately the same statistics as measurements on the state $|\psi(T)\rangle = U(T,0)|\psi(0)\rangle$. In particular, when an observable $Q$ with eigenvalues $\{q_1,q_2,\cdots\}$ is measured, the probabilities $P_{q_j}\big(|\psi(T)\rangle\big)$ and $P_{q_j}\big(|\psi_{\textrm{PF}(k)}(T)\rangle\big)$ of obtaining the eigenvalue $q_j$ when measuring $Q$ in the states $|\psi(T)\rangle$ and $|\psi_{\textrm{PF}(k)}(T)\rangle$, respectively, obey the relation \cite{Nielsen2010}
\begin{equation}
\big|P_{q_j}\big(|\psi(T)\rangle\big)-P_{q_j}\big(|\psi_{\textrm{PF}(k)}(T)\rangle\big)\big|\leq 2\epsilon_{\textrm{PF}(k)}(T,0).
\end{equation} 

\subsection{Classical optimization} \label{Sec:Classical optimization}

The control field optimization is accomplished iteratively using a classical optimization routine, which seeks to identify the set of control parameters that minimize $J[T,\{\theta_i\}]$. Global evolutionary strategies such as genetic algorithms have often been employed for this purpose \cite{Assion919,2002EPJD...20...71D,Levis709}. Gradient algorithms can also be used \cite{KHANEJA2005296}, although on quantum computers, obtaining the gradient information needed to implement these approaches requires additional measurements. Namely, if the value of $J[T,\{\theta_i\}]$ can be estimated in $\mathcal{O} (m)$ measurements, $\mathcal{O} (Km)$ additional measurements are required to estimate the $K$ gradients $\frac{\partial J[T,\{\theta_i\}]}{\partial \theta_j}$, $j=1,2,\cdots,K$, via finite differences. This increases the cost of optimization substantially per iteration compared to gradient-free algorithms\footnote{For some parametrized Hamiltonians it is possible to extract derivatives directly from measurements on the outputs of additional, related circuits \cite{schuld_evaluating_2018}. However, these approaches require Hamiltonians with particular structure that we do not assume, and moreover, they still require additional experiments.}. However, gradient algorithms may require fewer iterations to converge, suggesting that the choice of optimization method should be made considering the balance between measurement costs and classical optimization effort in mind. 

\section{Numerical illustrations} \label{Sec:Numerical illustrations}

In this section, we numerically investigate the performance of first-, second-, and fourth-order product formulas towards simulating the controlled dynamics of three model systems. For each system, the qubit encoding method used for our numerical analyses is described, with full details of the Hamiltonian and objective function mappings provided in Appendix \ref{Sec:Appendix}. We quantify the product formula performance by the Trotter error $\norm{U_{PF(k)}(T,0)-U(T,0)}$ and by the objective function error $|J_{PF(k)}-J|$, where $U(T,0)$ and $J$ denote the time evolution operator and objective function value in the numerically exact limit of $n\rightarrow\infty$, respectively, while $U_{PF(k)}(T,0)$ and $J_{PF(k)}$ are the corresponding values when a $k$-th order product formula is employed. In addition, possibilities for extensions towards more complex control applications are discussed for each case. We conclude this section with an analysis of the effects of hardware noise on the performance of the algorithm.

\subsection{Controlled bond stretching in HF} \label{Sec:Controlled bond stretching in diatomic molecule}

We first consider controlling the bond displacement of the diatomic molecule hydrogen fluoride (HF), modeled as a nonrotating Morse oscillator on the ground electronic state in the Born-Oppenheimer approximation \cite{PhysRev.34.57}. Morse oscillators have been used extensively in the literature as simple yet nontrivial proof-of-concept models for testing new quantum control ideas \cite{doi:10.1063/1.475576,PARAMONOV1993169,PhysRevA.37.4950,doi:10.1063/1.458438}. The drift Hamiltonian is given by
\begin{equation}
    H_0=\frac{p^2}{2m}  +V(r)\,,
    \label{MorseDrift}
\end{equation}
where $r$ is the bond coordinate operator, $p$ is the center of mass momentum operator, $m=1732\, m_e$ is the reduced mass of HF, and $V(r)$ is the anharmonic Morse potential
\begin{equation}
    V(r)= D(1-e^{-\alpha (r-r_0)})^2-D
    \label{MorsePotential}
\end{equation}
with equilibrium bond position $r_0 = 1.75\, a_0$, well depth $D=0.2101\, E_\text{h}$, and potential variation parameter $\alpha = 1.22\, a_0^{-1}$. We assume that the polarization of the field $f(t)$ is aligned with the system's dipole moment. Then, the control Hamiltonian in the dipole approximation is $H_c = -\mu(r)$, modeled by the function
\begin{equation}
    \mu(r) = \mu_0 re^{-\beta r^4}\,,
\end{equation}
where $\mu_0 = 0.4541\, \text{a.u.}$ specifies the strength of the dipole, and the parameter $\beta =0.0064\, a_0^{-4}$ governs the bond length of the maximum dipole moment \cite{GULDBERG1991229}. The form of this dipole moment function captures the fact that bonds of zero or infinite bond length do not contribute to the dipole moment, and it has two parameters which have been fitted to \emph{ab initio} data \cite{STINE1979161}.

We consider the task of driving the bond to a target bond length $\gamma= 1.5r_0$ at the terminal time $T$, and formulate the associated control objective function as
\begin{equation}
    J_{\mathrm{v}}[T,\{\theta_i\}]=\big(\langle \psi(T)|r|\psi(T)\rangle - \gamma\big)^2 \,.
    \label{TriatomicJ}
\end{equation}
This simple quantum control example can be extended to design fields for achieving controlled dissociation, \eg by setting the target bond length $\gamma$ to be sufficiently large. For dissociation, additional terms could also be added to $J_{\mathrm{v}}[T,\{\theta_i\}]$ that require the energy to be greater than the dissociation energy, or for the momentum to be positive (\ie such that the atoms are moving apart), at the terminal time $T$. The model could also be extended to simulate vibrational dynamics on multiple coupled electronic states, or to multiple coupled vibrational degrees of freedom, in order to simulate the vibrational dynamics of more complex systems.

To perform the qubit encoding, $H(t)$ is represented in a basis truncated to $d$ harmonic oscillator eigenfunctions, by evaluating its matrix elements in the harmonic oscillator basis via $\langle v| H(t)|v'\rangle = \int_{-\infty}^\infty v(\tilde{r})^*H(t,\tilde{r})v'(\tilde{r}) d\tilde{r}$, where $v(\tilde{r})$ are the harmonic oscillator eigenfunctions and we have introduced the shifted bond coordinate $\tilde{r}\equiv r-r_0$ which describes the displacement of the bond from its equilibrium position $r_0$ and centers the Morse potential at $\tilde{r}=0$. The resultant matrices are then expanded as weighted sums of Pauli basis operators as per Eq. (\ref{OperatorEncoding}). To perform the quantum dynamics simulation, the initial condition for the oscillator is set as the harmonic ground state (in a truncated basis of size $2^N$) $|\psi(0)\rangle = |0\rangle ^{\otimes N}$, which well-approximates the true ground state of the Morse oscillator and whose encoded qubit state is simple to prepare. The dynamics can be simulated using product formulas, and at the culmination of the quantum circuit, the qubits can be read out to determine $J_{\mathrm{v}}[T,\{\theta_i\}]$ by evaluating the expectation values of the $d\log_2(d)/2$ Pauli strings in the qubit operator encoding for $r$. The explicit operator encodings for $H_0$, $H_c$, and $r$ are given in Appendix \ref{Sec:Appendix}.

\begin{figure}
\includegraphics[width=1.0\columnwidth]{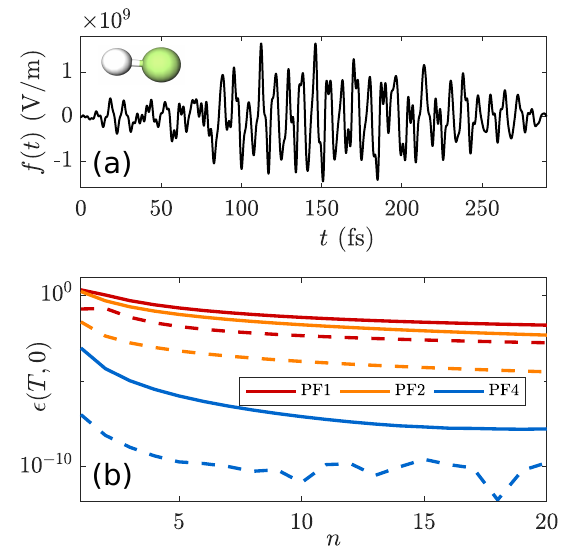}
\caption{In (a), the field $f(t)$ used for controlling the bond stretch of hydrogen fluoride with $J_{\mathrm{v}}[T,\{\theta_i\}]=0.01$ is shown, while in (b) the solid curves show the Trotter error for first-, second-, and fourth-order product formulas for different Trotter numbers $n$. The dashed curves show the corresponding objective function error, computed as $|J_{\mathrm{v},PF(k)}-J_{\mathrm{v}}|$.}
\label{FieldAndErrorFig}
\end{figure}

Fig.~\ref{FieldAndErrorFig}a shows a field that achieves $J_{\mathrm{v}}[T,\{\theta_i\}] = 0.01$ in the numerically exact $n\rightarrow\infty$ limit for an oscillator represented using $d=16$ harmonic oscillator eigenfunctions (\ie $N=4$ qubits), where $T = 290$ fs and $\Delta t = 0.024$ fs, parameterized as
\begin{equation}
    f(t,\{\theta_i\}) = \varepsilon(t)\bigg(\sum_{i} a_i\cos((\omega_i+\Delta_i)t-\phi_i)\bigg)\,,
    \label{MorseFieldEqn}
\end{equation}
where $\varepsilon(t)=\sin^{1/p}(\tfrac{\pi t}{T})$ is an envelope function, whose width is defined by the parameter $p$, and where $a_i$, $\Delta_i$, and $\phi_i$ denote the amplitude, detuning, and phase associated with the $i$-th frequency component, respectively. Optimal fields could be designed by optimizing over the set of control parameters $\{\theta_i\}=\{p,a_i,\Delta_i,\phi_i\}$. The solid curves in Fig.~\ref{FieldAndErrorFig}b show the total Trotter error $\epsilon(T,0)$ associated with using first-, second-, and fourth-order product formulas to simulate the field-induced dynamics from $t=0$ to $t=T$ as a function of the Trotter number $n$, while the dashed curves show the associated error in the objective function $J_{\mathrm{v}}[T,\{\theta_i\}]$. We remark that the Trotter error bounds given in Eqs. (\ref{epsilonPF1}) and (\ref{epsilonPF2p}) are multiple orders of magnitude greater than the actual Trotter error for this problem, and thus, are not plotted. The same is true for the examples presented in Sections \ref{Sec:Controlled orientation of two dipole-dipole coupled molecular rotors} and \ref{Sec:Controlled state preparation in light-harvesting complex}.  

\subsection{Controlled orientation of two dipole-dipole coupled OCS rotors} \label{Sec:Controlled orientation of two dipole-dipole coupled molecular rotors}

We next consider the problem of controlling the orientations of two dipole-dipole coupled carbonyl sulfide (OCS) molecules, modeled as linear rigid rotors in a plane. Experimentally, systems of planar molecular rotors could be formed by adsorbing cold molecules onto a surface or trapping them in an optical lattice, while shaped microwave control fields can be created experimentally with an arbitrary waveform generator \cite{Lin2005,Yao2011}. The controlled orientation of OCS molecules has been the subject of laboratory studies \cite{exp2,exp1} due to the importance of molecular orientation in applications including chemical reactions \cite{Brooks11,Zare1875,Rakitzis1852} and high harmonic generation \cite{PhysRevLett.109.233903}. Furthermore, the controlled orientation of dipole-dipole coupled OCS rotors using quantum optimal control has been studied theoretically in \cite{Yu2017,doi:10.1063/1.5091520}. 

\begin{figure*}
\includegraphics[width=1.85\columnwidth]{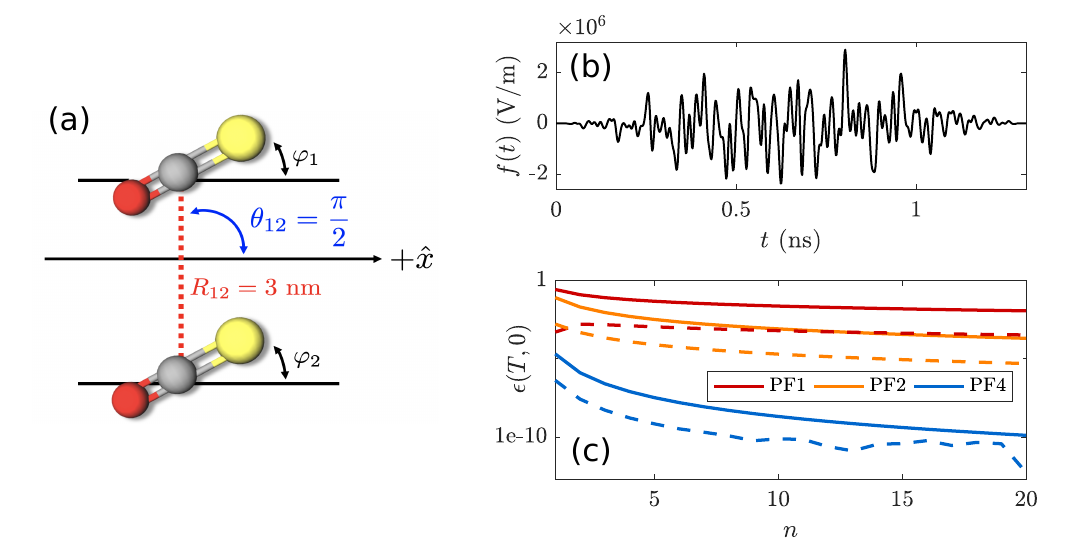}
\caption{In (a), a schematic diagram of the two dipole-dipole coupled planar OCS rotors is shown, where $\varphi_i$ is the angle of orientation of the $i$th rotor (relative to the $\hat{\bm{x}}$-axis), $\theta_{12}=\pi/2$ is the angle between the vector $\bm{R}_{12}$ between rotors and the $\hat{\bm{x}}$-axis, and $R_{12}=3$ nm is the distance between the two rotors. The field is assumed to be polarized along the $\hat{\bm{x}}$-axis. Figure (b) shows the microwave field $f(t)$ used for controlling the orientations of the two OCS rotors with $J_{\mathrm{r}}[T,\{\theta_i\}]=0.02$, in (c) the associated Trotter error $\epsilon(T,0)$ is plotted (solid curves) for first-, second-, and fourth-order product formulas for different Trotter numbers $n$. The objective function error, computed as $|J_{\mathrm{r},PF(k)}-J_{\mathrm{r}}|$, is also shown (dashed curves).} 
\label{RotFig}
\end{figure*}

The drift Hamiltonian of the coupled rotor system is given by
\begin{equation}
\begin{aligned}
H_0=\sum_{i=1}^2 H_i+V_{12}\,,
\label{eq:Hamiltonian}
\end{aligned}
\end{equation}
where the field-free, single-rotor Hamiltonian for the $i$th rotor is
\begin{equation}
\begin{aligned}
H_i=BL_i^2\, ,
\end{aligned}
\end{equation}
where $B=4.03\times 10^{-24}$ J is the rotational constant of OCS \cite{doi:10.1063/1.3253139} and $L_i^2 = -\hbar^2\frac{\partial^2}{\partial\varphi_i^2}$ is the squared angular momentum operator of rotor $i$, where $\varphi_i$ is the angular coordinate operator of the $i$-th rotor, and the angular coordinate represents the angle of the rotor's dipole moment with respect to the polarization direction of the field, assumed to be along the $\hat{\bm{x}}$-axis. The interaction describing the dipole-dipole coupling between the two rotors is 
\begin{equation}
\begin{aligned}
V_{12}&=\frac{\mu^2}{4\pi\epsilon_0 R_{12}^3}\Big\{(1-3\cos^2\theta_{12})\cos\varphi_1\cos\varphi_2 \\
&\quad +(1-3\sin^2\theta_{12})\sin\varphi_1\sin\varphi_2\\
&\quad -3\sin\theta_{12}\cos\theta_{12}(\cos\varphi_1\sin\varphi_2\\
&\quad +\sin\varphi_1\cos\varphi_2)\Big\}\,,
\end{aligned}
\end{equation}
where $\epsilon_0$ is the vacuum permittivity, $R_{12}=|\bm{R}_{12}|=3$ nm is the distance between the two rotors, and $\theta_{12}=\bm{R}_{12}\cdot\hat{\bm{x}}/R_{12}=\pi/2$ is the angle between the vector $\bm{R}_{12}$ between rotors and the $\hat{\bm{x}}$-axis. A schematic of the coupled rotor system is provided in Fig.~\ref{RotFig}a. 

The control Hamiltonian is given by
\begin{equation}
    H_c = -\mu(\cos\varphi_1+\cos\varphi_2)\,,
\end{equation}
where $\mu=2.36\times 10^{-30}$ C$\cdot$m is the magnitude of the permanent dipole moment of OCS \cite{PhysRev.77.500}. 

The minimization of the control objective function 
\begin{equation}
    J_{\mathrm{r}}[T,\{\theta_i\}]= 1-\tfrac{1}{\mathcal{N}_{\mathrm{r}}}\langle\psi(T)|(\cos\varphi_1+\cos\varphi_2)|\psi(T)\rangle
\end{equation}
then seeks both rotors to be identically oriented in the $+\hat{\bm{x}}$ direction at the terminal time $t=T$, where $\mathcal{N}_{\mathrm{r}} = \norm{\cos\varphi_1+\cos\varphi_2}$ is included for normalization and $\norm{\cdot}$ denotes the spectral norm.

The state of each rotor is represented using a truncated basis composed of tensor products of the eigenstates $|m_1\rangle$ of $L_1^2$ and $|m_2\rangle$ of $L_2^2$, where $m=-M,...,-1,0,1,\cdots,M$ and $M$ is set to 3, such that each rotor is represented using $d=2M+1 = 7$ levels, which can be encoded into $\lceil\log_2d\rceil = 3$ qubits. The eigenstates $|m_i\rangle$, $i=1,2$, satisfy the eigenvalue equation $L_i^2|m_i\rangle=m_i^2|m_i\rangle$ and can be expressed in terms of the angles $\varphi_i$ as $\langle\varphi_i|m_i\rangle=\sqrt{\frac{1}{2\pi}}e^{im_i\varphi_i}$, where $|\varphi_i\rangle$ are the eigenstates of the rotational coordinate operators. After representing $H_0$ and $H_c$ in this basis, the resultant matrices are expanded as weighted sums of Pauli basis operators (see Appendix \ref{Sec:Appendix}). Product formulas can be used to perform the quantum dynamics simulation using $N=6$ qubits, where each rotor is initialized in its ground state $|m_i=0\rangle\mapsto|1\rangle|0\rangle|0\rangle $ such that $|\psi(0)\rangle = |1\rangle|0\rangle|0\rangle \otimes |1\rangle|0\rangle|0\rangle $. At the culmination of the circuit, the qubits can be measured to determine $J_{\mathrm{r}}[T,\{\theta_i\}]$, whose explicit qubit encoding is given in Appendix \ref{Sec:Appendix}.  

Fig.~\ref{RotFig}b shows a control field that achieves $J_{\mathrm{r}}[T,\{\theta_i\}] = 0.02$ in the $n\rightarrow\infty$ limit for $T = 1.31$ ns and $\Delta t = 1.87$ ps. The field is taken to be a mix of ten frequency components with variable amplitudes $a_i$, detunings $\Delta_i$, and phases $\phi_i$, as per Eq. (\ref{MorseFieldEqn}). The solid curves in Fig.~\ref{RotFig}c show the Trotter error $\epsilon(T,0)$ associated with using first-, second-, and fourth-order product formulas to simulate the dynamics of the rotors from $t=0$ to $t=T$ as a function of the Trotter number $n$. The dashed curves show the associated objective function error.

We remark that extensions to systems consisting of $M>2$ coupled molecular rotors are straightforward, and the associated simulations can be performed using $N=M\lceil\log_2d\rceil$ qubits. Such simulations could be used to design optimally-shaped microwave fields to orient systems of numerous molecules with high precision, e.g., to improve the conversion of chemical reactions \cite{Stapelfeldt2003} or to improve the yield in high harmonic generation \cite{RevModPhys.80.117}.

\subsection{Controlled state preparation in light-harvesting complex} \label{Sec:Controlled state preparation in light-harvesting complex}

As a final example, we consider the task of excitonic state preparation in light-harvesting complexes. Understanding charge and energy flow in organic complexes such as photosynthetic light-harvesting complexes is a challenge due to the complexity of such systems and the confluence of several energy and timescales involved in these dynamical processes \cite{fleming_grand_2008,2007Natur.446..782E,Panitchayangkoon12766}. While advanced spectroscopic probes such as multidimensional spectroscopies \cite{schlau-cohen_two-dimensional_2011} provide insights into these dynamics, a challenge in such experiments is the preparation of localized initial excitonic states, such as those seen by these systems \emph{in vivo}. Previous research has studied the potential of quantum optimal control for preparing such initial states \cite{hoyer_realistic_2014}, and using shaped control pulses for controlling energy flow in light-harvesting complexes \cite{Herek2002QuantumCO, bruggemann_laser_2006}. 

Due to the complexity of light-harvesting dynamics, and the immense practical importance of understanding charge and energy flow in complex materials, this example proposes an important application of our algorithm for solving optimal control problems using quantum hardware. In this section we show through a simple example, how to map standard models of light-harvesting dynamics to a simulation circuit, and analyze the complexity of treating larger scale examples. 

\begin{figure}
\includegraphics[width=1.0\columnwidth]{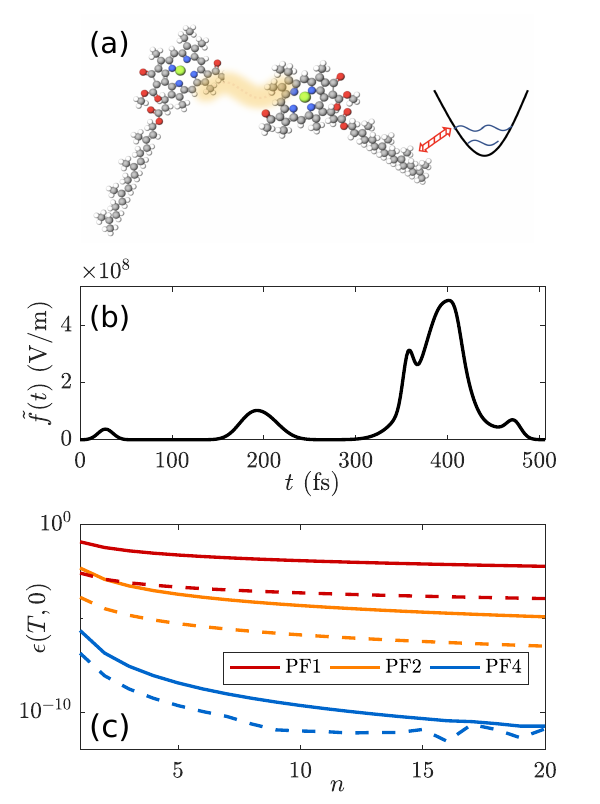}
\caption{(a) provides an illustration for the portion of the FMO light-harvesting complex of green sulfur bacteria considered here, which consists of two coupled chromophores, with one chromophore additionally coupled to a thermal vibrational mode. In (b), a field $f(t)$ that prepares an excitonic state on the second chromophore with $J_{\mathrm{s}}[T,\{\theta_i\}]=0.25$, is shown, while in (b) the Trotter error $\epsilon(T,0)$ is plotted (solid curves) for first-, second-, and fourth-order product formulas for different Trotter numbers $n$. The associated dashed curves show the objective function error, given by $|J_{\mathrm{s},PF(k)}-J_{\mathrm{s}}|$.}
\label{FieldAndError_Ill2}
\end{figure}

We consider a portion the Fenna-Matthews-Olson (FMO) complex of green sulfur bacteria. The structure of a full FMO monomer is known \cite{camara-artigas_structure_2003, ADOLPHS20062778, olbrich_atomistic_2011}, and consists of seven chromophores (bacteriochlorophyll A (BChl a) molecules) responsible for transferring energy towards the reaction center. The electronic excitations of the FMO complex couple to molecular vibrations and solvent degrees of freedom that are usually modeled as a Gaussian bosonic reservoir. However, recently it has become clear that certain modes of this reservoir are long-lived and moderately strongly coupled to the chromophores. This means that their coherent dynamics have a strong effect on the electronic excitation (exciton) dynamics, \eg \cite{womick_vibronic_2011, 2013NatPh...9..113C, irish_vibration-assisted_2014, dijkstra_coherent_2015, liu_vibrational_2016, thyrhaug_identification_2018}. Thus, simulating transport in complexes coupled to underdamped vibrations has become important for understanding the efficiency of energy transfer in such systems. As a minimal model of such a setting, we consider just chromophores 3 and 4 of FMO and the dominant vibrational mode at $180\text{ cm}^{-1}$ coupled to chromophore 4 \cite{2013NatPh...9..113C} (see Fig.~\ref{FieldAndError_Ill2}a). We ignore the other vibrational degrees of freedom for simplicity, which correspond to only capturing dynamics at short time-scales. In addition to the chromophores and vibrational mode, we model a global, weak electromagnetic field, $f(t)$, coupled to both chromophores. 

The drift Hamiltonian in this example is given by
\begin{equation}
\begin{aligned}
    H_0 =& \,\mathcal{E}_3b_3^\dagger b_3 + \mathcal{E}_4b_4^\dagger b_4 + \mathcal{J}_{3,4}(b_3^\dagger b_4 +
    b_4^\dagger b_3) + \nu a^\dagger a \\
    & + \mathcal{J}_{4,\upsilon}(b_4^\dagger b_4)(a+a^\dagger)\,,
    \label{Eq:LHCdrift}
    \end{aligned}
\end{equation}
where the subscripts label the FMO chromophores, $b_j$ denotes the annihilation operator for an electronic excitation on chromophore $j$, and $a$ and $a^\dagger$ are the harmonic oscillator lowering and raising operators, respectively. The parameters $\mathcal{E}_3 = 12,205\text{ cm}^{-1}$ and $\mathcal{E}_4 = 12,135\text{ cm}^{-1}$ denote the excitation energies on chromophores 3 and 4, while $\mathcal{J}_{3,4}=53.5\text{ cm}^{-1}$ denotes the dipole-dipole coupling between the two chromophores. We work in settings where the electromagnetic field is weak, and hence restrict the above Hamiltonian to at most one excitation per chromophore. The vibrational degree of freedom is modeled as a harmonic oscillator at thermal equilibrium at some temperature $T_{\text{vib}}$, where $\nu = 180\text{ cm}^{-1}$ denotes the frequency of the vibrational mode and $\mathcal{J}_{4,\upsilon} = 84.4\text{ cm}^{-1}$ denotes the magnitude of the coupling between chromophore 4 and the vibrational mode \cite{ADOLPHS20062778,2013NatPh...9..113C}. 

The control field is modeled as
\begin{equation}
    f(t,\{\theta_i\}) = \tilde{f}(t,\{\theta_i\})(e^{i\omega_0 t}+\textrm{c.c.})\,,
\end{equation}
where $\omega_0 = 12,200 \text{ cm}^{-1}$ is the carrier frequency and $\tilde{f}(t,\{\theta_i\})$ is the dimensionless field profile, whose shape can be optimized. This field is coupled to the system through the control Hamiltonian:
\begin{equation}
    H_c = \mu_3(b_3+b_3^\dagger)+\mu_4(b_4+b_4^\dagger)\,,
    \label{Eq:LHCcontrol}
\end{equation}
where $\mu_3 = 0.32|\mu|$ and $\mu_4 = 0.92|\mu|$ are the dipole couplings of each of the chromophores, $|\mu|=6.3$ Debye is the magnitude of the transition dipole for the relevant BChl a transition \cite{olbrich_atomistic_2011}, and the other factors account for the alignment of the chromophores with the polarization of the control field. We assume that chromophores 3 and 4 are oriented at angles $109^{\circ}$ and $23^{\circ}$, respectively, to the field's polarization direction \cite{Milder2010}. We simulate the coupled chromophore subsystem in a frame rotating at $\omega_0$ and make the rotating wave approximation as described in Ref.~\cite{hoyer_realistic_2014}, such that $\tilde{H}(t) = \tilde{H_0}+H_c\tilde{f}(t)$, where $\tilde{f}(t)$ is the field profile and in $\tilde{H_0}$, the excitation energies are shifted, $\tilde{\mathcal{E}}_3 =  \mathcal{E}_3 -\omega_0$ and $\tilde{\mathcal{E}}_4 = \mathcal{E}_4 -\omega_0$, while all other terms remain the same as in the original frame.

The coupled chromophore subsystem is initialized in the ground state
$|g_3g_4\rangle$. Meanwhile, the vibrational degree of freedom is taken to be
initially in the thermal state $\sum_{\upsilon = 0}^{\upsilon_{\max}} c_{\upsilon}(T_{\upsilon}, \upsilon_{\max}) |\upsilon\rangle\langle\upsilon|$,
with the coefficients given by
\begin{equation}
    c_{\upsilon}(T_{\upsilon}, \upsilon_{\max}) = \frac{1}{Z(T_{\upsilon}, \upsilon_{\max})} e^{-\beta(T_{\text{vib}})\, \varepsilon_{\upsilon}} ,
\end{equation}
where $Z(T_{\text{vib}}, \upsilon_{\max}) = \sum_{\upsilon}^{\upsilon_{\max}} e^{-\beta(T_{\text{vib}}) \varepsilon_{\upsilon}}$ is the partition function, $\varepsilon_{\upsilon} = \nu \upsilon$ is the energy associated with the $\upsilon$th eigenstate $|\upsilon\rangle$, and $\beta(T_{\text{vib}}) = \frac{1}{K_B T_{\text{vib}}}$ where $K_B$ is the Boltzmann constant, $T_{\text{vib}}=300 \text{ K}$ is the temperature, and we select $\upsilon_{\max}=7$, given that the higher vibrational states are not significantly occupied. The controlled preparation of an excited state localized on chromophore 4 can then be sought by minimizing
\begin{equation}
    J_{\mathrm{s}}[T,\{\theta_i\}] = \sum_{\upsilon = 0}^{\upsilon_{\max}} c_{\upsilon} (T_{\upsilon}, \upsilon_{\max}) J_{\upsilon}[T,\{\theta_i\}]\,,
    \label{Eq:ObjSum}
\end{equation} 
where 
\begin{equation}
    J_{\upsilon} [T,\{\theta_i\}]= 1-\langle \psi(T) |P\otimes I_{\upsilon} |\psi(T)\rangle,\,\,
    |\psi(0)\rangle = |g_3g_4\rangle| \upsilon \rangle \,,
    \label{Eq:Jv}
\end{equation}
where the projector $P \equiv |g_3e_4\rangle\langle g_3e_4|$, while $I_{\upsilon}$ denotes the identity operator on the vibrational degree of freedom. This objective function quantifies the state overlap of the chromophore subsystem with the target state $|g_3 e_4\rangle$ at time $T$, without specifying the final-time state of the vibrational degree of freedom. In order to evaluate Eq. (\ref{Eq:ObjSum}), a set of simulations can be performed with the vibrational mode initialized in each of the eigenstates $|\upsilon\rangle$, $\upsilon = 0, \cdots, \upsilon_{\max}$, as per Eq.~(\ref{Eq:Jv}). The control objective function $J_{\upsilon}[T,\{\theta_i\}]$ can be evaluated each time to determine the population in the target state $|g_3e_4\rangle$. Finally, the total objective can be calculated as a weighted sum of the results as per Eq. (\ref{Eq:ObjSum}).

The operators in $H_0$ and $H_c$ are mapped to qubit operators as $b_j^\dagger b_j\rightarrow -\sigma_{z,j}/2$, $(b_3^\dagger b_4+b_4^\dagger b_3)\rightarrow (\sigma_{+,3}\sigma_{-,4}+\sigma_{-,4}\sigma_{+,3})$, and $(b_j+b_j^\dagger)\rightarrow \sigma_{x,j}$, where $\sigma_+ = \bigl( \begin{smallmatrix}0 & 1\\ 0 & 0\end{smallmatrix}\bigr)$ and $\sigma_- = \bigl( \begin{smallmatrix}0 & 0\\ 1 & 0\end{smallmatrix}\bigr)$. To perform the qubit encoding for the vibrational subsystem, we represent it in a basis truncated to $d$ harmonic oscillator eigenfunctions, using the well-known matrix element relations for $a^\dagger a$ and $(a+a^\dagger)$, and expand the resultant matrices as weighted sums of Pauli basis operators (see Appendix \ref{Sec:Appendix}). 

Fig.~\ref{FieldAndError_Ill2}b shows a field that achieves $J_{\mathrm{s}}[T,\{\theta_i\}] = 0.25$ in the $n\rightarrow\infty$ limit, for $T = 508$ fs and $N_t = 300$, and Fig.~\ref{FieldAndError_Ill2}c shows the associated error $\epsilon(T,0)$ when first-, second-, and fourth-order product formulas are used. For the field plotted in Fig.~\ref{FieldAndError_Ill2}b, the field profile is taken to be a sum of ten Gaussian functions,
\begin{equation}
    \tilde{f}(t,\{\theta_i\}) = \varepsilon(t)\Big(\sum_{j=1}^{10} a_je^{-(t-b_jT)^2/(c_j T)^2}\Big)\,,
\end{equation} 
contained in the envelope $\varepsilon(t) = \sin^{1/2}(\tfrac{\pi t}{T})$, where the control parameters could be selected as $\{a_j\}$, $\{b_j\}$, and $\{c_j\}$, which govern the relative amplitudes, means, and variances of the Gaussian functions, respectively. 

\begin{figure*}
\includegraphics[width=2.08\columnwidth]{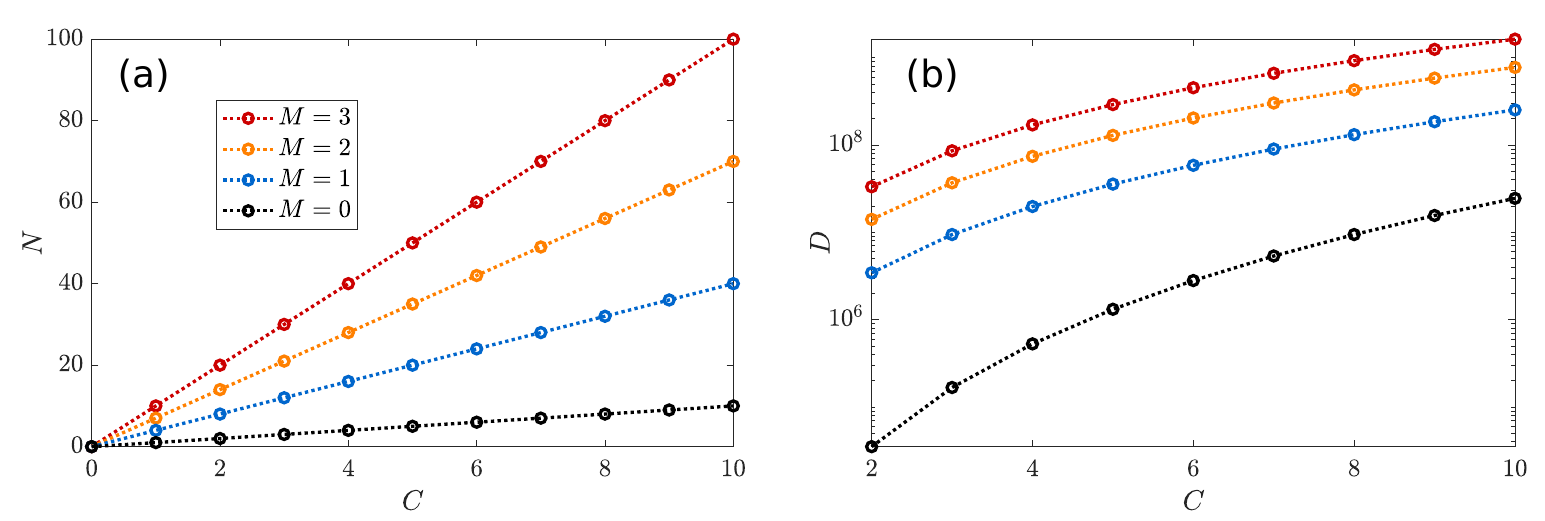}
\caption{(a) Required qubit counts, $N$, to simulate $C$ chromophores each coupled to $M$ vibrational modes, modeled using $d=8$ levels each. The qubit count for simulating an FMO model with 7 chromophores, and 2 modes per chromophore, is $N=49$. (b) An upper bound for the circuit depth $D$, quantified by the number of applications of $e^{-i B_\ell\tau_\ell}$, needed to simulate the dynamics over a single time-step $\Delta t$ using a fourth-order product formula for $C$ coupled chromophores each coupled to $M=0,1,2$ or $3$ vibrational modes modeled using $d=8$ levels. }
\label{ResourceAnalysisFig}
\end{figure*}

The procedure described here can be straightforwardly extended to quantum control simulations involving more complex models for light-harvesting complexes. Fig.~\ref{ResourceAnalysisFig}a shows the required qubit counts needed to simulate the dynamics of complexes composed of varying numbers of chromophores and vibrational modes. For example, to simulate the complete model for an FMO monomer involving seven chromophores, each coupled to two vibrational modes modeled using eight levels each, would require 49 qubits. In general, simulating $C$ chromophores with arbitrary dipole-dipole couplings, each coupled to $M$ vibrational modes modeled using $d$ levels, requires $N=(\log_2(d)M+1)C$ qubits. Fig.~\ref{ResourceAnalysisFig}b shows an upper bound for the circuit depth $D$, quantified by the number of applications of $e^{-i B_\ell\tau_\ell}$, for arbitrary $\tau_\ell$, required to simulate $C$ chromophores each coupled to $M=0,1,2$ or $3$ vibrational modes, modeled using $d=8$ levels each and simulated using a fourth-order product formula, over a single time step of length $\Delta t = 10$ a.u., presuming $\Lambda_{max} = 0.01$ and $L=(1+C+20M)C$, with an error threshold of $\epsilon_{PF4}(t+\Delta t,t) \leq10^{-5}$. 

\subsection{Effects of hardware noise on algorithm performance} \label{Sec:Effects of hardware noise on algorithm performance}

Until fault tolerant quantum computers are available, the implementation of quantum algorithms will be restricted to noisy, intermediate-scale quantum (NISQ) computing devices with nonzero error rates. As such, in this section we analyze the NISQ applicability of the hybrid quantum-classical scheme proposed here. We note that in this regard, very recent work \cite{castaldo2020quantum} has demonstrated that present-day quantum hardware based on superconducting circuits remains too noisy to implement product formula-based time-dependent quantum simulation algorithms, as studied here, due to prohibitively high error rates. 

In order to further study the NISQ-relevance of the hybrid quantum-classical algorithm proposed here, we perform a numerical analysis of its performance on quantum hardware with different nonzero error rates. In particular, we simulate the implementation of a simplified version of our example from Sec. \ref{Sec:Controlled state preparation in light-harvesting complex} on noisy hardware based on trapped ions. We consider the same drift and control Hamiltonians given above in Eqs. (\ref{Eq:LHCdrift}) and (\ref{Eq:LHCcontrol}), respectively. However, for this analysis, we take $\upsilon_{\max}=3$, and the control field profile $\tilde{f}(t,\{\theta_i\})$ is modeled as 
\begin{equation}
    \tilde{f}(t,\{\theta_i\}) =\varepsilon(t)\Big(\sin^2\big(\tfrac{2\pi t}{T} - \phi_a\big)+\sin^2\big(\tfrac{2\pi t}{T} - \phi_b\big)\Big) \,,
    \label{Eq:fsimple}
\end{equation} 
where $T = 169$ fs, $\varepsilon(t) = 7.5\times 10^{-4} \sin(\frac{\pi t}{T})$ is an envelope function, and $\phi_a$ and $\phi_b$ are the control parameters. 

The system is initialized as before, with both chromophores in their respective ground states, and with the vibrational mode in a thermal state. Then, the controlled preparation of an excited state localized on chromophore 4 is sought by minimizing Eq. (\ref{Eq:ObjSum}) as before, i.e., by computing the full objective $J_s$ as a weighted average of the results from a set of simulations where the vibrational mode initialized in each of the eigenstates $|\upsilon\rangle$, $\upsilon = 0, 1,2,3$, as per Eq.~(\ref{Eq:Jv}). We remark that due to the form of Eq. (\ref{Eq:fsimple}), the objective function is symmetric with respect to $\phi_a$ and $\phi_b$, and extrema occur when $\phi_a$ and $\phi_b$ are chosen such that they constructively interfere, i.e., when $\phi_a=\phi_b$, or when $\phi_a-\phi_b$ is an integer multiple of $\pi$.

In order to simulate the implementation of this model on quantum hardware, the qubit encoding step is carried out as before. In this simplified setting, $N=4$ qubits are required to represent the system. Then, a first order product formula is used to simulate the field-induced dynamics over $N_t = 28$ time steps, using a Trotter number of $n=1$. This simulation is first compiled into a quantum circuit formed by a gate set consisting single-qubit rotations and CNOT gates, as per the recipe given in Fig. \ref{CircuitSchematic}. Then, this circuit is recompiled into a circuit formed by the native gate set associated with the trapped ion hardware model adapted from \cite{trout2018simulating}, containing $R_x(\pm\pi/2)$, $R_y(\pm\pi/2)$, $R_z(\theta)$ for arbitrary $\theta$, and the two-qubit M{\o}lmer-S{\o}rensen gate. In order to simulate error in the implementation of the native gates, all single-qubit  $R_x(\pm\pi/2)$ and $R_y(\pm\pi/2)$ gates are followed by a series of three quantum channels: an imprecise rotation about the rotation axis $\alpha = x,y$, 
\begin{equation}
\mathcal{E}_{\alpha,p_\alpha}(\rho) = (1-p_\alpha)\rho+p_\alpha R_\alpha(\pi)\rho R_\alpha(\pi)\,,
\end{equation}
a depolarizing channel, 
\begin{equation}
\mathcal{D}_{p_{\text{dep}}}(\rho) = (1-p_{\text{dep}})\rho+\frac{p_{\text{dep}}}{3}\big(\sigma_x \rho \sigma_x+\sigma_y\rho \sigma_y + \sigma_z\rho \sigma_z\big)\, ,
\end{equation}
and a dephasing channel,
\begin{equation}
\mathcal{Z}_{p_{\text{d}}}(\rho) = (1-p_{\text{d}})\rho+p_{\text{d}}\sigma_z\rho \sigma_z.
\end{equation}
The $R_z(\theta)$ gates are assumed to be implemented virtually and incur no error.
Meanwhile, the two-qubit M{\o}lmer-S{\o}rensen gate is followed by an imprecise rotation about the $\sigma_x \sigma_x$ axis,
\begin{equation}
\mathcal{E}_{2,p_{\text{xx}}}(\rho) = (1-p_{\text{xx}})\rho+p_{\text{xx}}\sigma_x\sigma_x\rho \sigma_x \sigma_x\, ,
\end{equation}
a motional heating error (i.e., an imprecise rotation with a different error rate), 
\begin{equation}
\mathcal{E}_{2,p_{\text{h}}}(\rho) = (1-p_{\text{h}})\rho+p_{\text{h}}\sigma_x \sigma_x\rho \sigma_x \sigma_x\, ,
\end{equation}
independent depolarization channels on each of the two qubits, i.e., $\mathcal{D}_{p_{\text{dep}}}\otimes \mathcal{D}_{p_{\text{dep}}} (\rho)$, and independent dephasing channels on each of the two qubits, i.e., $\mathcal{Z}_{p_{\text{d,2}}}\otimes \mathcal{Z}_{p_{\text{d,2}}}(\rho)$. In addition, each idle gate is replaced with a depolarizing channel with depolarization error rate $p_{\text{idle}} = p_{\text{dep}}/10$. Meanwhile, for state preparation, an ideal preparation of the $|0\rangle$ state is followed by a depolarizing channel with the same depolarization error rate $p_{\text{dep}}$. For measurement, an ideal measurement in the computational $\sigma_z$ basis of any qubit is preceded by a depolarizing channel, with the depolarization error rate equal to the largest imprecise rotation error rate (for further model details, see \cite{trout2018simulating}).

\begin{table}[]
\begin{tabular}{|c|c|c|} 
\hline
                 & \textbf{Realistic}   & \textbf{Optimistic} \\ \hline
$p_x$            & $10^{-4}$            & $10^{-4}$           \\ \hline
$p_y$            & $10^{-4}$            & $10^{-4}$           \\ \hline
$p_{\text{dep}}$ & $8\times10^{-4}$     & $10^{-6}$           \\ \hline
$p_{\text{xx}}$  & $10^{-3}$            & $10^{-3}$           \\ \hline
$p_\text{h}$     & $1.25\times 10^{-3}$ & $5\times 10^{-5}$   \\ \hline
$p_\text{d}$     & $1.5\times 10^{-4}$  & $10^{-5}$           \\ \hline
$p_{\text{d,2}}$ & $7.5\times 10^{-4}$  & $5\times 10^{-5}$   \\ \hline
\end{tabular}
\caption{Realistic and optimistic error rates associated with an error model for trapped-ion quantum hardware adapted from \cite{trout2018simulating}.}
\label{ErrorRateTable}
\end{table}

Within this noisy framework, we examine the algorithm performance using realistic error rates, optimistic error rates, and zero error rates. The realistic error rates are taken from \cite{trout2018simulating}, while the optimistic error rates are estimated based on projected improvements in ion heating rates and reductions in photon scattering (Raman scattering) while performing single and two qubit qubit gates. These error rates are collected in Table \ref{ErrorRateTable}.

For the noise-free, optimistic, and realistic cases, we then incorporate one additional source of error: the effect of using $m$ measurement samples to estimate the objective over the course of the control parameter optimization, where $m$ is a finite number (here, we consider both $m=10^3$ and $m=10^5$). We denote the objective that is found with a first order product formula, using this trapped-ion hardware model, by $J_{s,PF1}^{(H)}$, and reserve the use of $J_s$ to denote the ideal objective function, as per Sec. \ref{Sec:Controlled state preparation in light-harvesting complex}. 

\begin{figure}
\includegraphics[width=1.0\columnwidth]{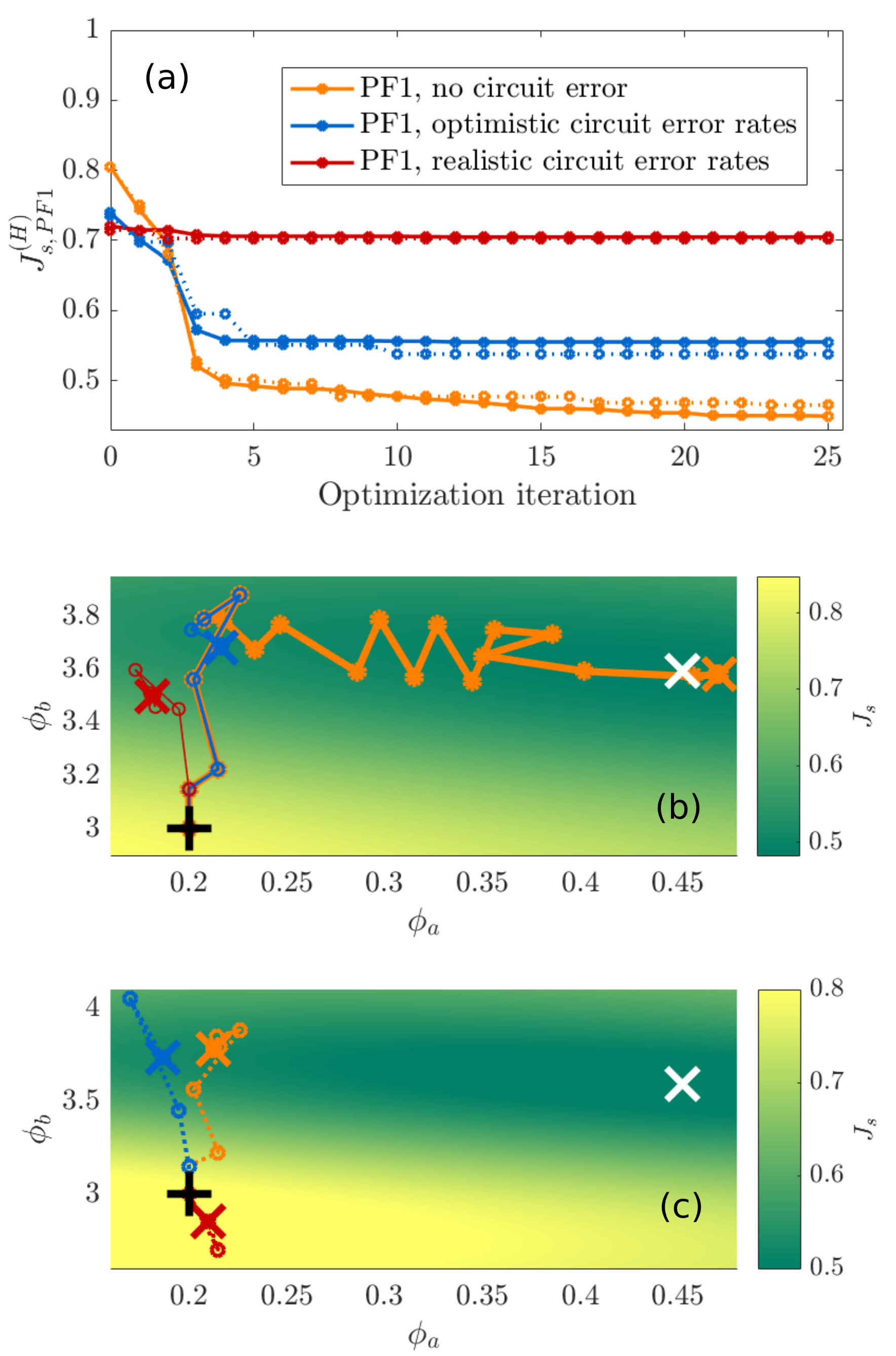}
\caption{(a) The value of $J_{s,PF1}^{(H)}$ evaluated using a model for trapped-ion hardware is shown for 25 optimization iterations. The value of $J_{s,PF1}^{(H)}$ is computed using a quantum circuit based on a first order product formula with no error (orange), optimistic error rates (red), and realistic error rates (red). Solid (dotted) curves correspond to a scenario where $m=10^5$ ($m=10^3$) measurement samples are used to estimate all expectation values required for the optimization. (b) The optimization trajectories associated with the case of $m=10^5$, with no circuit error (orange), optimistic circuit error rates (blue), and realistic circuit error rates (red) are shown superimposed on a map of the ideal optimization landscape (i.e., the landscape associated with a perfect simulation containing no circuit error, $n=\infty$, and $m=\infty$). All trajectories begin at the black `+', and terminate in an `x', with the minimum point marked by the white 'x'. Panel (c) shows analogous results for the case of $m=10^3$.}
\label{HeatMaps}
\end{figure}

We simulate the performance of the algorithm introduced in Sec. \ref{Sec:Hybrid algorithm for quantum optimal control} in these different settings. In order to optimize $J_{s,PF1}^{(H)}$, a Nelder-Mead algorithm is used, with control parameters initialized asymmetrically, as $\phi_a=0.2$ and $\phi_b=3$. The results of these numerical studies are given in Fig. \ref{HeatMaps}. In panel (a), the value of $J_{s,PF1}^{(H)}$ is shown as a function of optimization iteration for each of the three noise levels considered, where the solid curves correspond to $m=10^5$ and the dotted curves correspond to $m=10^3$. Meanwhile, panels (b) and (c) show the optimization trajectories plotted against the ideal landscape of $J_s$ (i.e., associated with $m,n\rightarrow\infty$).

For the case of no circuit error, shown in orange, the only errors present are due to the use of a first-order product formula to simulate the dynamics, and the use of a finite number of measurements $m$ to estimate the objective function at each iteration. As such, the associated optimization trajectory terminates very near to the true minimum for $m=10^5$, as shown in panel (b), and also relatively close for $m=10^3$ as shown in panel (c), and the optimization proceeds in the correct direction with respect to $\phi_a$ and $\phi_b$ for both $m$ values. When optimistic error rates are considered, the optimization trajectory terminates farther from the true minimum, but the optimization proceeds nevertheless in the correct direction with respect to $\phi_b$ for both $m$ values. However, for $m=10^3$, it moves in the wrong direction with respect to $\phi_a$. Finally, when realistic error rates are used, we see that the performance deteriorates much more, and the optimization proceeds in the wrong direction with respect to $\phi_a$ for $m=10^5$ and in the wrong direction with respect to $\phi_b$ for $m=10^3$. 

We also remark that the quantitative differences between $J_{s,PF1}^{(H)}$ and $J_{s}$ for the cases involving realistic error rates are significant. Namely, the terminal value of $J_{s,PF1}^{(H)}$ for both values of $m$ (i.e., associated with terminal $\{\phi_a,\phi_b\}$ values of $\{0.18,3.50\}$ and $\{0.21,2.85\}$) is $J_{s,PF1}^{(H)} = 0.70$. This is significantly different from the corresponding values of $J_s$, which are $J_s = 0.58$ for $\{\phi_a,\phi_b\} = \{0.18,3.50\}$ and $J_s = 0.83$ for $\{\phi_a,\phi_b\} = \{0.21,2.85\}$. The differences between $J_{s,PF1}^{(H)}$ and $J_{s}$ for the optimistic and error-free cases are less significant (i.e., the terminal values of $J_s$ are within 10\% of the associated $J_{s,PF1}^{(H)}$ values).

These findings suggest that realistic trapped-ion error rates are too significant for a quantitatively-accurate implementation of the needed quantum circuits. However, as hardware improves, we can expect improvements in the quality of the simulations that are possible, as indicated by the positive trends in Fig. \ref{HeatMaps} as the error rate is reduced from realistic to optimistic to error-free. Furthermore, the results may also be further improved if methods such as readout error mitigation and zero noise extrapolation are used as in \cite{fauseweh2020digital}.

\section{Discussion and Outlook} \label{Sec:Outlook}

In this article we have explored how quantum computers could be used for simulations of molecular control, which can often be intractable on classical computers. We introduced an algorithm that utilizes a quantum computer to simulate the field-induced dynamics of the molecular system under consideration, and a classical co-processor to optimize a set of control field parameters to achieve a desired objective.  Three numerical illustrations were then presented; the first two examples considered vibrational and rotational control problems, while the third treated the problem of state preparation in light-harvesting complexes, which could serve as a potential benchmark problem. To this end,  in Fig \ref{ResourceAnalysisFig}, we analyzed the qubit counts and circuit depths required for its solution on a quantum computer. 

One key difference between our approach and most other variational quantum algorithms is that the solution is encoded in the variational control parameters $\{\theta_i\}$, rather than the terminal state of the qubits. It should also be noted that unlike most other variational quantum algorithms, which are designed to use shallow quantum circuits compatible with noisy quantum devices, the depth of the quantum circuits associated with our algorithm can vary arbitrarily depending on factors such as the pulse length, time-step size, and the desired error tolerance.  

Given this situation, we considered the prospects of a near-term implementation of our algorithm on NISQ hardware. In particular, we provided a numerical analysis of its performance in the presence of different levels of hardware noise, using a model for trapped ion hardware, with results presented in Fig. \ref{HeatMaps}. Although we found that quantitatively-accurate calculations on NISQ hardware are likely infeasible with current hardware error rates, we wish to emphasize that there may exist important application settings where precise quantitative accuracy is not required. For example, in settings where the molecular systems under consideration become more complex, the study of trends may become increasingly important. Such situations are prevalent across the field of chemistry, where systematic theoretical, computational, and experimental studies allow for exploring trends governing the reactivity of chemical reagents in different conditions. In this spirit, an emerging goal in the domain of quantum control is to gain an understanding of the trends associated with controlled molecular dynamics. In this latter setting, control fields are akin to \emph{photonic reagents} \cite{rabitz2003shaped}, and it is insights into their role in driving chemical processes that are sought \cite{moore2013systematic,tibbetts2013optimal}. Consequently, future studies using our algorithm to explore the systematics of molecular control may be useful in revealing important relationships and trends between the properties of control fields, and the dynamical behavior they induce in chemical systems. Such studies may tolerate certain levels of hardware noise as long as the correct trends remain present, thereby suggesting a potential application area of our algorithm in the NISQ era. 
 
Furthermore, beyond the illustrative examples treated explicitly in this article, numerous additional applications of molecular control can be imagined. Many of these will surely involve more complex systems than the simple models we consider, such as polyatomic molecules with interacting vibrational, rotational, and electronic degrees of freedom, whose controlled dynamics can involve a breakdown of the Born-Oppenheimer approximation. It is expected that real space encodings outside of the Born-Oppenheimer approximation may be useful for simulating such systems \cite{Kassal18681}. In real space, the state of an electron or nucleus represented using $d$ points on a 1D grid can be encoded in a binary fashion using $\lceil \log_2 d\rceil$ qubits, by treating each grid point analogously to how each basis function was treated in Sec. \ref{Sec:Qubit encoding}. This means that a total of $N = 3M\lceil \log_2 d\rceil$ qubits would needed to simulate the controlled dynamics of a polyatomic molecule composed of $M$ interacting electrons and nuclei represented this way. Simulations of molecular dynamics performed in this manner require computational resources far beyond the capabilities of today's classical computers. However, if such simulations could be performed on quantum hardware in the future, they could enable the identification of optimal fields for controlling the outcome of chemical reactions, which is of fundamental interest, and also of potential utility towards commercially relevant chemical synthesis applications. In the emerging area of attosecond control, such simulations could also be used to explore the possibility of designing fields to control charge-directed reactivity, where attosecond pulses are sought to control the electron dynamics such that the associated nuclei are driven towards a desired chemical reaction path \cite{2014NaPho...8..195L}.

\begin{acknowledgments}

A.B.M. acknowledges helpful discussions with W.J. Huggins and C. Arenz, and support from the U.S. Department of Energy, Office of Science, Office of Advanced Scientific Computing Research, Department of Energy Computational Science Graduate Fellowship under Award No. DE-FG02-97ER25308. H.A.R. acknowledges support from NSF Grant No. CHE-1763198.

A.B.M. was supported by the U.S. Department of Energy, Office of Science, Office of Advanced Scientific Computing Research, under the Quantum Computing Application Teams program.
M.D.G. was supported by the U.S. Department of Energy, Office of Science, Office of Advanced Scientific Computing Research, under the Quantum Computing Application Teams program and Quantum Algorithm Teams programs.
M.S. was supported by the U.S. Department of Energy, Office of Science,  Office of Basic Energy Sciences under the Materials and Chemical Sciences Research for Quantum Information Research program.

This report was prepared as an account of work sponsored by an agency of the United States Government. Neither the United States Government nor any agency thereof, nor any of their employees, makes any warranty, express or implied, or assumes any legal liability or responsibility for the accuracy, completeness, or usefulness of any information, apparatus, product, or process disclosed, or represents that its use would not infringe privately owned rights. Reference herein to any specific commercial product, process, or service by trade name, trademark, manufacturer, or otherwise does not necessarily constitute or imply its endorsement, recommendation, or favoring by the United States Government or any agency thereof. The views and opinions of authors expressed herein do not necessarily state or reflect those of the United States Government or any agency thereof.

Sandia National Laboratories is a multimission laboratory managed and operated by National Technology \& Engineering Solutions of Sandia, LLC, a wholly owned subsidiary of Honeywell International Inc., for the U.S. Department of Energy's National Nuclear Security Administration under contract DE-NA0003525. This paper describes objective technical results and analysis. Any subjective views or opinions that might be expressed in the paper do not necessarily represent the views of the U.S. Department of Energy or the United States Government. 
\end{acknowledgments}

\bibliography{bib2.bib}

\appendix

\section{Qubit operator encodings} \label{Sec:Appendix}

In this appendix, the full details regarding the qubit operator encodings are given for each of the numerical illustrations presented in the main text. The mappings for the Hamiltonian are performed as,
\begin{equation}
\begin{aligned}
H(t) &= H_0+H_cf(t)\\
&\mapsto \sum_{\ell=1}^L g_{0,\ell} B_\ell +\sum_{\ell=1}^L g_{c,\ell} B_\ell f(t)\\
& = \sum_{\ell=1}^L g_\ell (t) B_\ell
\end{aligned}
\end{equation}
where $B_\ell$ are normalized tensor products of Pauli operators (here denoted using  $X=\frac{1}{\sqrt{2}} \sigma_x$, $Y=\frac{1}{\sqrt{2}} \sigma_y$, $Z =\frac{1}{\sqrt{2}}  \sigma_z$, and $I$, where $I$ is the $2\times 2$ identity matrix, normalized using a prefactor of $\frac{1}{\sqrt{2}} $), and where the time-dependent coefficients 
\begin{equation}
g_\ell (t) = g_{0,\ell}+g_{c,\ell} f(t)
\end{equation}
and where $g_{0,\ell} =\langle H_0, B_\ell\rangle_{\textrm{HS}}$ and $g_{c,\ell}=\langle H_c, B_\ell\rangle_{\textrm{HS}}$ for $\ell = 1,\cdots,L$ and $H_0, H_c$ in atomic units.

The encoding for any operator $A$ in the control objective function $J[T,\{\theta_i\}]$ whose expectation value is sought is then given by,
\begin{equation}
\begin{aligned}
A &\mapsto \sum_{\ell} J_{\ell} B_\ell 
\end{aligned}
\end{equation}
where $J_{\ell} =\langle A, B_\ell\rangle_{\textrm{HS}}$ .

\begin{table*}[]
\begin{tabular}{|c|c|c|l|c|c|c|lccc}
\cline{1-3} \cline{5-7} \cline{9-11}
$B_\ell$    & $g_{0,\ell}$           & $g_{c.\ell}$           &  & $B_\ell$    & $g_{0,\ell}$           & $g_{c.\ell}$            & \multicolumn{1}{l|}{} & \multicolumn{1}{c|}{$B_\ell$}    & \multicolumn{1}{c|}{$g_{0,\ell}$ }           & \multicolumn{1}{c|}{$g_{c.\ell}$ }           \\ \cline{1-3} \cline{5-7} \cline{9-11} 
XXXX & -0.038560085 & 0.05114078   &  & YYIZ & -0.001268035 & 2.028E-05    & \multicolumn{1}{l|}{} & \multicolumn{1}{c|}{IXXX} & \multicolumn{1}{c|}{-0.089739199} & \multicolumn{1}{c|}{0.097560999}  \\ \cline{1-3} \cline{5-7} \cline{9-11} 
XXXZ & -0.002351416 & 0.001110148  &  & YYII & 0.009446845  & -0.000187368 & \multicolumn{1}{l|}{} & \multicolumn{1}{c|}{IXXZ} & \multicolumn{1}{c|}{-0.0049006}   & \multicolumn{1}{c|}{0.002219332}  \\ \cline{1-3} \cline{5-7} \cline{9-11} 
XXXI & 0.017912988  & -0.017352443 &  & YZXY & 0.000533084  & -1.98106E-05 & \multicolumn{1}{l|}{} & \multicolumn{1}{c|}{IXXI} & \multicolumn{1}{c|}{0.046719828}  & \multicolumn{1}{c|}{-0.034819353} \\ \cline{1-3} \cline{5-7} \cline{9-11} 
XXYY & 0.017808426  & -0.053815291 &  & YZYX & 0.000558687  & -1.9942E-05  & \multicolumn{1}{l|}{} & \multicolumn{1}{c|}{IXYY} & \multicolumn{1}{c|}{0.040197373}  & \multicolumn{1}{c|}{-0.103275856} \\ \cline{1-3} \cline{5-7} \cline{9-11} 
XXZX & 0.004579774  & 0.000482563  &  & YZYZ & 2.47182E-05  & 7.79482E-07  & \multicolumn{1}{l|}{} & \multicolumn{1}{c|}{IXZX} & \multicolumn{1}{c|}{0.00939786}   & \multicolumn{1}{c|}{0.000918422}  \\ \cline{1-3} \cline{5-7} \cline{9-11} 
XXZZ & 0.000193455  & -1.33605E-06 &  & YZYI & -0.00018603  & -6.53938E-06 & \multicolumn{1}{l|}{} & \multicolumn{1}{c|}{IXZZ} & \multicolumn{1}{c|}{0.000389917}  & \multicolumn{1}{c|}{-2.73222E-06} \\ \cline{1-3} \cline{5-7} \cline{9-11} 
XXZI & -0.002531285 & 4.07023E-05  &  & YZZY & -8.38841E-06 & 3.14905E-08  & \multicolumn{1}{l|}{} & \multicolumn{1}{c|}{IXZI} & \multicolumn{1}{c|}{-0.00537161}  & \multicolumn{1}{c|}{7.23422E-05}  \\ \cline{1-3} \cline{5-7} \cline{9-11} 
XXIX & -0.021001708 & -0.002685675 &  & YZIY & 2.64627E-05  & -1.32288E-07 & \multicolumn{1}{l|}{} & \multicolumn{1}{c|}{IXIX} & \multicolumn{1}{c|}{-0.050058066} & \multicolumn{1}{c|}{-0.005740823} \\ \cline{1-3} \cline{5-7} \cline{9-11} 
XXIZ & -0.001268035 & 2.02799E-05  &  & YIXY & -0.000892729 & 3.40134E-05  & \multicolumn{1}{l|}{} & \multicolumn{1}{c|}{IXIZ} & \multicolumn{1}{c|}{-0.002691003} & \multicolumn{1}{c|}{3.60407E-05}  \\ \cline{1-3} \cline{5-7} \cline{9-11} 
XXII & 0.009446845  & -0.000187368 &  & YIYX & -0.000926616 & 3.42018E-05  & \multicolumn{1}{l|}{} & \multicolumn{1}{c|}{IXII} & \multicolumn{1}{c|}{0.02509691}   & \multicolumn{1}{c|}{-0.00041776}  \\ \cline{1-3} \cline{5-7} \cline{9-11} 
XYXY & 0.018591474  & -0.053845549 &  & YIYZ & -4.88208E-05 & -1.68703E-06 & \multicolumn{1}{l|}{} & \multicolumn{1}{c|}{IYXY} & \multicolumn{1}{c|}{-0.042638967} & \multicolumn{1}{c|}{0.103366352}  \\ \cline{1-3} \cline{5-7} \cline{9-11} 
XYYX & 0.037751015  & -0.05111038  &  & YIYI & 0.000273994  & 9.97501E-06  & \multicolumn{1}{l|}{} & \multicolumn{1}{c|}{IYYX} & \multicolumn{1}{c|}{-0.087177061} & \multicolumn{1}{c|}{0.097469993}  \\ \cline{1-3} \cline{5-7} \cline{9-11} 
XYYZ & 0.002306284  & -0.001111752 &  & YIZY & 1.40903E-05  & -6.70093E-08 & \multicolumn{1}{l|}{} & \multicolumn{1}{c|}{IYYZ} & \multicolumn{1}{c|}{-0.00478065}  & \multicolumn{1}{c|}{0.002223208}  \\ \cline{1-3} \cline{5-7} \cline{9-11} 
XYYI & -0.017688312 & 0.017360857  &  & YIIY & -3.60324E-05 & 1.95879E-07  & \multicolumn{1}{l|}{} & \multicolumn{1}{c|}{IYYI} & \multicolumn{1}{c|}{0.045872962}  & \multicolumn{1}{c|}{-0.034848607} \\ \cline{1-3} \cline{5-7} \cline{9-11} 
XYZY & -0.004039326 & -0.000502317 &  & ZXXX & 0.064913944  & -0.021409988 & \multicolumn{1}{l|}{} & \multicolumn{1}{c|}{IYZY} & \multicolumn{1}{c|}{0.00814128}   & \multicolumn{1}{c|}{0.000961914}  \\ \cline{1-3} \cline{5-7} \cline{9-11} 
XYIY & 0.019352792  & 0.002748225  &  & ZXXZ & 0.00272657   & -2.68863E-05 & \multicolumn{1}{l|}{} & \multicolumn{1}{c|}{IYIY} & \multicolumn{1}{c|}{-0.044940194} & \multicolumn{1}{c|}{-0.005925961} \\ \cline{1-3} \cline{5-7} \cline{9-11} 
XZXX & 0.000559009  & -1.99323E-05 &  & ZXXI & -0.038166502 & 0.017757373  & \multicolumn{1}{l|}{} & \multicolumn{1}{c|}{IZXX} & \multicolumn{1}{c|}{0.063941015}  & \multicolumn{1}{c|}{-0.024662828} \\ \cline{1-3} \cline{5-7} \cline{9-11} 
XZXZ & 2.47336E-05  & 7.79388E-07  &  & ZXYY & -0.02797267  & 0.025213657  & \multicolumn{1}{l|}{} & \multicolumn{1}{c|}{IZXZ} & \multicolumn{1}{c|}{0.002689118}  & \multicolumn{1}{c|}{-0.00000196}  \\ \cline{1-3} \cline{5-7} \cline{9-11} 
XZXI & -0.000186089 & -6.53895E-06 &  & ZXZX & -0.00404338  & -0.000188292 & \multicolumn{1}{l|}{} & \multicolumn{1}{c|}{IZXI} & \multicolumn{1}{c|}{-0.038862822} & \multicolumn{1}{c|}{0.01779779}   \\ \cline{1-3} \cline{5-7} \cline{9-11} 
XZYY & -0.000532767 & 1.98204E-05  &  & ZXZZ & -0.000078083 & -2.24322E-06 & \multicolumn{1}{l|}{} & \multicolumn{1}{c|}{IZYY} & \multicolumn{1}{c|}{0.028833035}  & \multicolumn{1}{c|}{-0.028409772} \\ \cline{1-3} \cline{5-7} \cline{9-11} 
XZZX & -8.68401E-06 & 2.28843E-08  &  & ZXZI & 0.00310499   & -2.15828E-05 & \multicolumn{1}{l|}{} & \multicolumn{1}{c|}{IZZX} & \multicolumn{1}{c|}{-0.009134215} & \multicolumn{1}{c|}{-0.0082323}   \\ \cline{1-3} \cline{5-7} \cline{9-11} 
XZZZ & -3.2453E-07  & 7.2097E-09   &  & ZXIX & 0.036998894  & 0.003791397  & \multicolumn{1}{l|}{} & \multicolumn{1}{c|}{IZZZ} & \multicolumn{1}{c|}{-0.000121}    & \multicolumn{1}{c|}{-0.0000035}   \\ \cline{1-3} \cline{5-7} \cline{9-11} 
XZZI & 2.44708E-06  & -5.69297E-08 &  & ZXIZ & 0.001552597  & -1.07923E-05 & \multicolumn{1}{l|}{} & \multicolumn{1}{c|}{IZZI} & \multicolumn{1}{c|}{0.0078275}    & \multicolumn{1}{c|}{-0.000117}    \\ \cline{1-3} \cline{5-7} \cline{9-11} 
XZIX & 2.71493E-05  & -1.1145E-07  &  & ZXII & -0.02071289  & 0.000312001  & \multicolumn{1}{l|}{} & \multicolumn{1}{c|}{IZIX} & \multicolumn{1}{c|}{0.092660185}  & \multicolumn{1}{c|}{-0.0547794}   \\ \cline{1-3} \cline{5-7} \cline{9-11} 
XZIZ & 1.23554E-06  & -2.86952E-08 &  & ZYXY & 0.03018479   & -0.025294895 & \multicolumn{1}{l|}{} & \multicolumn{1}{c|}{IZIZ} & \multicolumn{1}{c|}{0.0039145}    & \multicolumn{1}{c|}{-0.000059}    \\ \cline{1-3} \cline{5-7} \cline{9-11} 
XZII & -6.58044E-06 & 1.56164E-07  &  & ZYYX & 0.062583996  & -0.02132826  & \multicolumn{1}{l|}{} & \multicolumn{1}{c|}{IZII} & \multicolumn{1}{c|}{-0.20927511}  & \multicolumn{1}{c|}{0.0350575}    \\ \cline{1-3} \cline{5-7} \cline{9-11} 
XIXX & -0.000926984 & 3.41906E-05  &  & ZYYZ & 0.00262638   & -2.99737E-05 & \multicolumn{1}{l|}{} & \multicolumn{1}{c|}{IIXX} & \multicolumn{1}{c|}{-0.183273085} & \multicolumn{1}{c|}{0.191318632}  \\ \cline{1-3} \cline{5-7} \cline{9-11} 
XIXZ & -4.88394E-05 & -1.6869E-06  &  & ZYYI & -0.037362908 & 0.017784867  & \multicolumn{1}{l|}{} & \multicolumn{1}{c|}{IIXZ} & \multicolumn{1}{c|}{-0.009773262} & \multicolumn{1}{c|}{0.00446384}   \\ \cline{1-3} \cline{5-7} \cline{9-11} 
XIXI & 0.00027406   & 9.97454E-06  &  & ZYZY & -0.00313268  & -0.000218172 & \multicolumn{1}{l|}{} & \multicolumn{1}{c|}{IIXI} & \multicolumn{1}{c|}{0.097970198}  & \multicolumn{1}{c|}{-0.06968541}  \\ \cline{1-3} \cline{5-7} \cline{9-11} 
XIYY & 0.000892365  & -3.40248E-05 &  & ZYIY & 0.032412446  & 0.003955379  & \multicolumn{1}{l|}{} & \multicolumn{1}{c|}{IIYY} & \multicolumn{1}{c|}{-0.085518865} & \multicolumn{1}{c|}{0.203152768}  \\ \cline{1-3} \cline{5-7} \cline{9-11} 
XIZX & 1.44817E-05  & -5.53019E-08 &  & ZZXX & -0.024611285 & -0.010947388 & \multicolumn{1}{l|}{} & \multicolumn{1}{c|}{IIZX} & \multicolumn{1}{c|}{0.046233285}  & \multicolumn{1}{c|}{-0.028844}    \\ \cline{1-3} \cline{5-7} \cline{9-11} 
XIZZ & 6.4178E-07   & -1.48099E-08 &  & ZZXZ & -0.000366282 & 0.00002124   & \multicolumn{1}{l|}{} & \multicolumn{1}{c|}{IIZZ} & \multicolumn{1}{c|}{0.0019585}    & \multicolumn{1}{c|}{-0.0000295}   \\ \cline{1-3} \cline{5-7} \cline{9-11} 
XIZI & -3.61417E-06 & 8.52905E-08  &  & ZZXI & 0.021424378  & -0.00019621  & \multicolumn{1}{l|}{} & \multicolumn{1}{c|}{IIZI} & \multicolumn{1}{c|}{-0.104758555} & \multicolumn{1}{c|}{0.017525}     \\ \cline{1-3} \cline{5-7} \cline{9-11} 
XIIX & -3.68441E-05 & 1.70953E-07  &  & ZZYY & -0.010228865 & -0.010127212 & \multicolumn{1}{l|}{} & \multicolumn{1}{c|}{IIIX} & \multicolumn{1}{c|}{-0.271137115} & \multicolumn{1}{c|}{0.3917319}    \\ \cline{1-3} \cline{5-7} \cline{9-11} 
XIIZ & -1.84801E-06 & 4.35472E-08  &  & ZZZX & -0.000181115 & -0.0057793   & \multicolumn{1}{l|}{} & \multicolumn{1}{c|}{IIIZ} & \multicolumn{1}{c|}{-0.052394277} & \multicolumn{1}{c|}{0.008762}     \\ \cline{1-3} \cline{5-7} \cline{9-11} 
XIII & 8.33357E-06  & -1.98986E-07 &  & ZZZZ & 0.0000105    & -0.0000005   &                       &                           &                                   &                                   \\ \cline{1-3} \cline{5-7}
YXXY & -0.018591585 & 0.053845546  &  & ZZZI & -0.000965    & -0.000032    &                       &                           &                                   &                                   \\ \cline{1-3} \cline{5-7}
YXYX & -0.037751127 & 0.051110376  &  & ZZIX & -0.035030315 & -0.0226364   &                       &                           &                                   &                                   \\ \cline{1-3} \cline{5-7}
YXYZ & -0.00230629  & 0.001111752  &  & ZZIZ & -0.000482    & -0.000016    &                       &                           &                                   &                                   \\ \cline{1-3} \cline{5-7}
YXYI & 0.01768833   & -0.017360857 &  & ZZII & 0.0312035    & -0.0004655   &                       &                           &                                   &                                   \\ \cline{1-3} \cline{5-7}
YXZY & 0.004039328  & 0.000502317  &  & ZIXX & 0.127593015  & -0.045175368 &                       &                           &                                   &                                   \\ \cline{1-3} \cline{5-7}
YXIY & -0.019352794 & -0.002748225 &  & ZIXZ & 0.005361338  & -0.00003316  &                       &                           &                                   &                                   \\ \cline{1-3} \cline{5-7}
YYXX & -0.038559973 & 0.051140784  &  & ZIXI & -0.076261802 & 0.03557119   &                       &                           &                                   &                                   \\ \cline{1-3} \cline{5-7}
YYXZ & -0.00235141  & 0.001110148  &  & ZIYY & 0.057968835  & -0.052851232 &                       &                           &                                   &                                   \\ \cline{1-3} \cline{5-7}
YYXI & 0.01791297   & -0.017352443 &  & ZIZX & -0.017689615 & -0.012631    &                       &                           &                                   &                                   \\ \cline{1-3} \cline{5-7}
YYYY & 0.017808315  & -0.053815294 &  & ZIZZ & -0.000241    & -0.0000075   &                       &                           &                                   &                                   \\ \cline{1-3} \cline{5-7}
YYZX & 0.004579772  & 0.000482563  &  & ZIZI & 0.0156125    & -0.000233    &                       &                           &                                   &                                   \\ \cline{1-3} \cline{5-7}
YYZZ & 0.000193455  & -1.33605E-06 &  & ZIIX & 0.185460385  & -0.1001331   &                       &                           &                                   &                                   \\ \cline{1-3} \cline{5-7}
YYZI & -0.002531285 & 4.07024E-05  &  & ZIIZ & 0.0078075    & -0.000116    &                       &                           &                                   &                                   \\ \cline{1-3} \cline{5-7}
YYIX & -0.021001706 & -0.002685675 &  & ZIII & -0.416622219 & 0.0701785    &                       &                           &                                   &                                   \\ \cline{1-3} \cline{5-7}
\end{tabular}
\caption{Hamiltonian encoding for Sec. \ref{Sec:Controlled bond stretching in diatomic molecule}: Controlled bond stretching in HF. Table lists basis operators $B_\ell$, $\ell = 1,\cdots,135$ and corresponding nonzero coefficients $g_{0,\ell}$ and $g_{c,\ell}$ associated with encoding the Hamiltonian of the Morse oscillator in the Pauli operator basis.}
\end{table*}

\begin{table}[]
\begin{tabular}{|c|c|}
\hline
$B_\ell$   & $J_\ell$         \\ \hline
XXXX & 1.414213562  \\ \hline
XXYY & -1.414213562 \\ \hline
XYXY & -1.414213562 \\ \hline
XYYX & -1.414213562 \\ \hline
YXXY & 1.414213562  \\ \hline
YXYX & 1.414213562  \\ \hline
YYXX & 1.414213562  \\ \hline
YYYY & -1.414213562 \\ \hline
ZXXX & -0.732050808 \\ \hline
ZXYY & 0.732050808  \\ \hline
ZYXY & -0.732050808 \\ \hline
ZYYX & -0.732050808 \\ \hline
ZZXX & -0.227948227 \\ \hline
ZZYY & -0.227948227 \\ \hline
ZZZX & -0.136587377 \\ \hline
ZZIX & -0.493929325 \\ \hline
ZIXX & -1.520115871 \\ \hline
ZIYY & -1.520115871 \\ \hline
ZIZX & -0.27883864  \\ \hline
ZIIX & -3.090644658 \\ \hline
IXXX & 2.732050808  \\ \hline
IXYY & -2.732050808 \\ \hline
IYXY & 2.732050808  \\ \hline
IYYX & 2.732050808  \\ \hline
IZXX & -0.807327954 \\ \hline
IZYY & -0.807327954 \\ \hline
IZZX & -0.185780097 \\ \hline
IZIX & -1.655839156 \\ \hline
IIXX & 5.383819176  \\ \hline
IIYY & 5.383819176  \\ \hline
IIZX & -0.862895501 \\ \hline
IIIX & 10.70451475  \\ \hline
\end{tabular}
\caption{Objective function encoding for Sec. \ref{Sec:Controlled bond stretching in diatomic molecule}: Controlled bond stretching in HF. Table lists basis operators $B_\ell$, $\ell = 1,\cdots,32$ and corresponding nonzero coefficients $J_{\ell}$ associated with encoding the bond coordinate operator $r$ in the Pauli operator basis, as needed for evaluating the objective function.}
\end{table}

\begin{table}[]
\begin{tabular}{ccclllllllllllllll}
\cline{1-3}
\multicolumn{1}{|c|}{$B_\ell$} & \multicolumn{1}{c|}{$g_{0,\ell}$ (e-4)} & \multicolumn{1}{c|}{$g_{c,\ell}$ (e-11)} &  &  &  &  &  &  &  &  &  &  &  &  &  &  &  \\ \cline{1-3}
\multicolumn{1}{|c|}{XXXIII}     & \multicolumn{1}{c|}{0}                 & \multicolumn{1}{c|}{-0.054252182}       &  &  &  &  &  &  &  &  &  &  &  &  &  &  &  \\ \cline{1-3}
\multicolumn{1}{|c|}{XYYIII}     & \multicolumn{1}{c|}{0}                 & \multicolumn{1}{c|}{0.054252182}        &  &  &  &  &  &  &  &  &  &  &  &  &  &  &  \\ \cline{1-3}
\multicolumn{1}{|c|}{YXYIII}     & \multicolumn{1}{c|}{0}                 & \multicolumn{1}{c|}{-0.054252182}       &  &  &  &  &  &  &  &  &  &  &  &  &  &  &  \\ \cline{1-3}
\multicolumn{1}{|c|}{YYXIII}     & \multicolumn{1}{c|}{0}                 & \multicolumn{1}{c|}{-0.054252182}       &  &  &  &  &  &  &  &  &  &  &  &  &  &  &  \\ \cline{1-3}
\multicolumn{1}{|c|}{ZZXIII}     & \multicolumn{1}{c|}{0}                 & \multicolumn{1}{c|}{0.054252182}        &  &  &  &  &  &  &  &  &  &  &  &  &  &  &  \\ \cline{1-3}
\multicolumn{1}{|c|}{ZZZIII}     & \multicolumn{1}{c|}{0.148027232}       & \multicolumn{1}{c|}{0}                  &  &  &  &  &  &  &  &  &  &  &  &  &  &  &  \\ \cline{1-3}
\multicolumn{1}{|c|}{ZZIIII}     & \multicolumn{1}{c|}{0.148027232}       & \multicolumn{1}{c|}{0}                  &  &  &  &  &  &  &  &  &  &  &  &  &  &  &  \\ \cline{1-3}
\multicolumn{1}{|c|}{ZIXIII}     & \multicolumn{1}{c|}{0}                 & \multicolumn{1}{c|}{-0.054252182}       &  &  &  &  &  &  &  &  &  &  &  &  &  &  &  \\ \cline{1-3}
\multicolumn{1}{|c|}{IXXIII}     & \multicolumn{1}{c|}{0}                 & \multicolumn{1}{c|}{-0.108504365}       &  &  &  &  &  &  &  &  &  &  &  &  &  &  &  \\ \cline{1-3}
\multicolumn{1}{|c|}{IYYIII}     & \multicolumn{1}{c|}{0}                 & \multicolumn{1}{c|}{-0.108504365}       &  &  &  &  &  &  &  &  &  &  &  &  &  &  &  \\ \cline{1-3}
\multicolumn{1}{|c|}{IZXIII}     & \multicolumn{1}{c|}{0}                 & \multicolumn{1}{c|}{-0.054252182}       &  &  &  &  &  &  &  &  &  &  &  &  &  &  &  \\ \cline{1-3}
\multicolumn{1}{|c|}{IZZIII}     & \multicolumn{1}{c|}{-0.074013616}      & \multicolumn{1}{c|}{0}                  &  &  &  &  &  &  &  &  &  &  &  &  &  &  &  \\ \cline{1-3}
\multicolumn{1}{|c|}{IZIIII}     & \multicolumn{1}{c|}{0.074013616}       & \multicolumn{1}{c|}{0}                  &  &  &  &  &  &  &  &  &  &  &  &  &  &  &  \\ \cline{1-3}
\multicolumn{1}{|c|}{IIXIII}     & \multicolumn{1}{c|}{0}                 & \multicolumn{1}{c|}{-0.162756547}       &  &  &  &  &  &  &  &  &  &  &  &  &  &  &  \\ \cline{1-3}
\multicolumn{1}{|c|}{IIZIII}     & \multicolumn{1}{c|}{0.111020424}       & \multicolumn{1}{c|}{0}                  &  &  &  &  &  &  &  &  &  &  &  &  &  &  &  \\ \cline{1-3}
\multicolumn{1}{|c|}{IIIXXX}     & \multicolumn{1}{c|}{0}                 & \multicolumn{1}{c|}{-0.054252182}       &  &  &  &  &  &  &  &  &  &  &  &  &  &  &  \\ \cline{1-3}
\multicolumn{1}{|c|}{IIIXYY}     & \multicolumn{1}{c|}{0}                 & \multicolumn{1}{c|}{0.054252182}        &  &  &  &  &  &  &  &  &  &  &  &  &  &  &  \\ \cline{1-3}
\multicolumn{1}{|c|}{IIIYXY}     & \multicolumn{1}{c|}{0}                 & \multicolumn{1}{c|}{-0.054252182}       &  &  &  &  &  &  &  &  &  &  &  &  &  &  &  \\ \cline{1-3}
\multicolumn{1}{|c|}{IIIYYX}     & \multicolumn{1}{c|}{0}                 & \multicolumn{1}{c|}{-0.054252182}       &  &  &  &  &  &  &  &  &  &  &  &  &  &  &  \\ \cline{1-3}
\multicolumn{1}{|c|}{IIIZZX}     & \multicolumn{1}{c|}{0}                 & \multicolumn{1}{c|}{0.054252182}        &  &  &  &  &  &  &  &  &  &  &  &  &  &  &  \\ \cline{1-3}
\multicolumn{1}{|c|}{IIIZZZ}     & \multicolumn{1}{c|}{0.148027232}       & \multicolumn{1}{c|}{0}                  &  &  &  &  &  &  &  &  &  &  &  &  &  &  &  \\ \cline{1-3}
\multicolumn{1}{|c|}{IIIZZI}     & \multicolumn{1}{c|}{0.148027232}       & \multicolumn{1}{c|}{0}                  &  &  &  &  &  &  &  &  &  &  &  &  &  &  &  \\ \cline{1-3}
\multicolumn{1}{|c|}{IIIZIX}     & \multicolumn{1}{c|}{0}                 & \multicolumn{1}{c|}{-0.054252182}       &  &  &  &  &  &  &  &  &  &  &  &  &  &  &  \\ \cline{1-3}
\multicolumn{1}{|c|}{IIIIXX}     & \multicolumn{1}{c|}{0}                 & \multicolumn{1}{c|}{-0.108504365}       &  &  &  &  &  &  &  &  &  &  &  &  &  &  &  \\ \cline{1-3}
\multicolumn{1}{|c|}{IIIIYY}     & \multicolumn{1}{c|}{0}                 & \multicolumn{1}{c|}{-0.108504365}       &  &  &  &  &  &  &  &  &  &  &  &  &  &  &  \\ \cline{1-3}
\multicolumn{1}{|c|}{IIIIZX}     & \multicolumn{1}{c|}{0}                 & \multicolumn{1}{c|}{-0.054252182}       &  &  &  &  &  &  &  &  &  &  &  &  &  &  &  \\ \cline{1-3}
\multicolumn{1}{|c|}{IIIIZZ}     & \multicolumn{1}{c|}{-0.074013616}      & \multicolumn{1}{c|}{0}                  &  &  &  &  &  &  &  &  &  &  &  &  &  &  &  \\ \cline{1-3}
\multicolumn{1}{|c|}{IIIIZI}     & \multicolumn{1}{c|}{0.074013616}       & \multicolumn{1}{c|}{0}                  &  &  &  &  &  &  &  &  &  &  &  &  &  &  &  \\ \cline{1-3}
\multicolumn{1}{|c|}{IIIIIX}     & \multicolumn{1}{c|}{0}                 & \multicolumn{1}{c|}{-0.162756547}       &  &  &  &  &  &  &  &  &  &  &  &  &  &  &  \\ \cline{1-3}
\multicolumn{1}{|c|}{IIIIIZ}     & \multicolumn{1}{c|}{0.111020424}       & \multicolumn{1}{c|}{0}                  &  &  &  &  &  &  &  &  &  &  &  &  &  &  &  \\ \cline{1-3}
\end{tabular}
\caption{Hamiltonian encodings for Sec. \ref{Sec:Controlled orientation of two dipole-dipole coupled molecular rotors}: Controlled orientation of two dipole-dipole coupled OCS rotors. Table lists basis operators $B_\ell$, $\ell = 1,\cdots,30$ and corresponding nonzero coefficients $g_{0,\ell}$ and $g_{c,\ell}$ associated with encoding the Hamiltonian of the dipole-dipole coupled rotors in the Pauli operator basis.}
\end{table}

\begin{table}[]
\begin{tabular}{|c|c|l}
\cline{1-2}
$B_\ell$     & $J_\ell$ &  \\ \cline{1-2}
XXXIII & 1  &  \\ \cline{1-2}
XYYIII & -1 &  \\ \cline{1-2}
YXYIII & 1  &  \\ \cline{1-2}
YYXIII & 1  &  \\ \cline{1-2}
ZZXIII & -1 &  \\ \cline{1-2}
ZIXIII & 1  &  \\ \cline{1-2}
IXXIII & 2  &  \\ \cline{1-2}
IYYIII & 2  &  \\ \cline{1-2}
IZXIII & 1  &  \\ \cline{1-2}
IIXIII & 3  &  \\ \cline{1-2}
IIIXXX & 1  &  \\ \cline{1-2}
IIIXYY & -1 &  \\ \cline{1-2}
IIIYXY & 1  &  \\ \cline{1-2}
IIIYYX & 1  &  \\ \cline{1-2}
IIIZZX & -1 &  \\ \cline{1-2}
IIIZIX & 1  &  \\ \cline{1-2}
IIIIXX & 2  &  \\ \cline{1-2}
IIIIYY & 2  &  \\ \cline{1-2}
IIIIZX & 1  &  \\ \cline{1-2}
IIIIIX & 3  &  \\ \cline{1-2}
\end{tabular}
\caption{Objective function encodings for Sec. \ref{Sec:Controlled orientation of two dipole-dipole coupled molecular rotors}: Controlled orientation of two dipole-dipole coupled OCS rotors. Table lists basis operators $B_\ell$, $\ell = 1,\cdots,20$ and corresponding nonzero coefficients $J_{\ell}$ associated with encoding the operator $\cos\varphi_1+\cos\varphi_2$ in the Pauli operator basis, as needed for evaluating the objective function.}
\end{table}

\begin{table}[]
\begin{tabular}{|c|c|c|}
\hline
$B_\ell$   & $g_{0,\ell}$          & $g_{c,\ell}$         \\ \hline
XXIII & 0.000689463  & 0           \\ \hline
XIIII & 0            & 4.489849672 \\ \hline
YYIII & 0.000689463  & 0           \\ \hline
ZIIII & -6.44358E-05 & 0           \\ \hline
IXIII & 0            & 12.90831781 \\ \hline
IZXXX & -0.000543838 & 0           \\ \hline
IZXYY & 0.000543838  & 0           \\ \hline
IZYXY & -0.000543838 & 0           \\ \hline
IZYYX & -0.000543838 & 0           \\ \hline
IZZXX & 0.000281511  & 0           \\ \hline
IZZYY & 0.000281511  & 0           \\ \hline
IZZZX & 8.76579E-05  & 0           \\ \hline
IZZIX & 0.000584563  & 0           \\ \hline
IZIXX & -0.001050615 & 0           \\ \hline
IZIYY & -0.001050615 & 0           \\ \hline
IZIZX & 0.000310459  & 0           \\ \hline
IZIIX & -0.002070357 & 0           \\ \hline
IZIII & -0.001739767 & 0           \\ \hline
IIZII & -0.009278757 & 0           \\ \hline
IIIZI & -0.004639379 & 0           \\ \hline
IIIIZ & -0.002319689 & 0           \\ \hline
\end{tabular}
\caption{Hamiltonian encodings for Sec. \ref{Sec:Controlled state preparation in light-harvesting complex}: Controlled state preparation in light-harvesting complex. Table lists basis operators $B_\ell$, $\ell = 1,\cdots,21$ and corresponding nonzero coefficients $g_{0,\ell}$ and $g_{c,\ell}$ associated with encoding the Hamiltonian of the model of a portion of the FMO complex of green sulfur bacteria in the Pauli operator basis.}
\end{table}

\begin{table}
\begin{tabular}{|c|c|}
\hline
$B_\ell$    & $J_\ell$     \\ \hline
ZZIII & -1.4142 \\ \hline
ZIIII & 1.4142  \\ \hline
IZIII & -1.4142 \\ \hline
IIIII & 1.4142  \\ \hline
\end{tabular}
\caption{Objective function encodings for Sec. \ref{Sec:Controlled state preparation in light-harvesting complex}: Controlled state preparation in light-harvesting complex. Table lists basis operators $B_\ell$, $\ell = 1,\cdots,4$ and corresponding nonzero coefficients $J_{\ell}$ associated with encoding the projector $P\otimes I_{\upsilon}$ in the Pauli operator basis, as needed for evaluating the objective function.}
\end{table}

\begin{table}[]
\begin{tabular}{|c|c|c|}
\hline
$B_\ell$   & $g_{0,\ell}$          & $g_{c,\ell}$         \\ \hline
XXII & 0.000487524 & 0           \\ \hline
XIII & 0            & 3.174803 \\ \hline
YYII & 0.000487524  & 0           \\ \hline
ZIII & -0.000045563 & 0           \\ \hline
IXII & 0            & 9.127559 \\ \hline
IZXX & -0.000543838 & 0           \\ \hline
IZXY & 0.000543838  & 0           \\ \hline
IZZX & 0.000281511 & 0           \\ \hline
IZIX & -0.00105061 & 0           \\ \hline
IZII & -0.001230201  & 0           \\ \hline
IIZI & -0.003280536 & 0           \\ \hline
IIIZ & -0.001640268  & 0           \\ \hline
\end{tabular}
\caption{Hamiltonian encodings for Sec. \ref{Sec:Effects of hardware noise on algorithm performance}: Effects of hardware noise on algorithm performance. Table lists basis operators $B_\ell$, $\ell = 1,\cdots,12$ and corresponding nonzero coefficients $g_{0,\ell}$ and $g_{c,\ell}$ associated with encoding the Hamiltonian of the simplified model of a portion of the FMO complex of green sulfur bacteria in the Pauli operator basis.}
\end{table}

\begin{table}
\begin{tabular}{|c|c|}
\hline
$B_\ell$    & $J_\ell$     \\ \hline
ZZII & -1.4142 \\ \hline
ZIII & 1.4142  \\ \hline
IZII & -1.4142 \\ \hline
IIII & 1.4142  \\ \hline
\end{tabular}
\caption{Objective function encodings for Sec. \ref{Sec:Effects of hardware noise on algorithm performance}: Effects of hardware noise on algorithm performance. Table lists basis operators $B_\ell$, $\ell = 1,\cdots,4$ and corresponding nonzero coefficients $J_{\ell}$ associated with encoding the projector $P\otimes I_{\upsilon}$ in the Pauli operator basis, as needed for evaluating the objective function.}
\end{table}

\end{document}